\documentclass[aps, prd, superscriptaddress, preprintnumbers, floatfix, longbibliography, twocolumn,10pt]{revtex4-2}

\usepackage{graphicx,amsfonts,amsmath,amssymb,amstext}
\usepackage{float,wrapfig}
\usepackage{subfigure, psfrag}
\usepackage{dsfont}
\usepackage{xcolor}
\usepackage[colorlinks=true,urlcolor=blue,citecolor=blue,linkcolor=blue]{hyperref}
\usepackage{bm}

\newcommand{\be}{\begin{equation}}
\newcommand{\ee}{\end{equation}}
\newcommand{\ba}{\begin{eqnarray}}
\newcommand{\ea}{\end{eqnarray}}

\definecolor{purple}{rgb}{0.8,0,0.6}
\definecolor{darkgreen}{rgb}{0.00,0.6,0.00}

\extrafloats{100}

\begin{document}

\title{Anomalous Gurzhi effect}
\date{February 23, 2021}

\author{P.~O.~Sukhachov}
\email{pavlo.sukhachov@yale.edu}
\affiliation{Department of Physics, Yale University, New Haven, Connecticut 06520, USA}

\author{Bj\"{o}rn Trauzettel}
\affiliation{Institut f\"{u}r Theoretische Physik und Astrophysik, Universit\"{a}t W\"{u}rzburg, 97074 W\"{u}rzburg, Germany}
\affiliation{W\"{u}rzburg-Dresden Cluster of Excellence ct.qmat, Germany}

\begin{abstract}
We investigate the role of the chiral anomaly in hydrodynamic and crossover regimes of transport in a Weyl or Dirac semimetal film. We show that the magnetic-field-dependent part of the electric conductivity in the direction of the magnetic field develops an unusual nonmonotonic dependence on temperature dubbed the anomalous Gurzhi effect. This effect is realized in sufficiently clean semimetals subject to a classically weak magnetic field where the electron-electron scattering dominates and the relaxation of the valley-imbalance charge density happens at the boundaries. Moreover, the conditions for the realization of the conventional Gurzhi effect in hydrodynamic and crossover regimes of transport in three-dimensional Dirac and Weyl semimetals are determined.
\end{abstract}

\maketitle

\section{Introduction}
\label{sec:Introduction}

Dirac and Weyl semimetals represent a unique class of crystals where quasiparticles have a relativistic-like linear dispersion relation in the vicinity of band-touching points called Dirac points and Weyl nodes~\cite{Burkov:rev-2018,Armitage:rev-2018,GMSS:book}. To describe the properties of such materials, it is imperative to use the corresponding relativistic-like equations, which resemble those in quantum electrodynamics. This allows us to realize several exotic effects in solid-state setups. We focus on one such effect, namely, the chiral anomaly~\cite{Adler:1969,BJ:1969}.

The chiral anomaly appears as the breakdown of the classical chiral symmetry via quantum effects. In the absence of electromagnetic fields, the numbers of left- and right-handed Weyl fermions are conserved separately. However, as was shown by Adler, Bell, and Jackiw~\cite{Adler:1969,BJ:1969}, the conservation law of chiral (imbalance) charge breaks down due to the chiral anomaly activated by electric and magnetic fields. The chiral anomaly in a condensed matter setting was for the first time probed in superfluid ${}^3$He~\cite{Bevan-Volovik:1997}; see, e.g., Ref.~\cite{Volovik:book-2003} for details of chiral anomaly manifestations in liquid helium. As for solids, it was predicted in Ref.~\cite{Nielsen:1983} that the chiral anomaly can affect the conductivity of materials with a linear dispersion relation in the vicinity of the band-touching points that are now known as Weyl semimetals. The chiral anomaly activated by electric and magnetic fields leads to the pumping of chiral quasiparticles between the Weyl nodes of opposite chiralities which results in accumulation of the chiral imbalance (chiral charge) density. This allows for the chiral magnetic effect current~\cite{Vilenkin:1980,Fukushima:2008} and, as a result, leads to the decrease of the resistivity by the magnetic field known as the ``negative" magnetoresistivity phenomenon. Negative magnetoresistivity was observed in Weyl (transition metal monopnictides TaAs, NbAs, TaP, and NbP) and Dirac (Na$_3$Bi, Cd$_3$As$_2$, and ZrTe$_5$) semimetals (see Refs.~\cite{Hosur-Qi:rev-2013,Burkov:rev-2015,Gorbar:2017lnp,Hu-Mao:rev-2019,Ong-Liang:rev-2020} for reviews on anomalous transport properties).

Among several transport regimes, the hydrodynamic one has recently received significant attention. It is realized when the electron-electron interactions dominate over the scattering of electrons on impurities and phonons. The hydrodynamic regime of charge and heat transport in solids was proposed a long time ago in the 1960s~\cite{Gurzhi:1963,Gurzhi:1968,Nielsen-Shklosvkii:1969a}. The first experimental signatures of hydrodynamic transport were observed only three decades later in the 1990s in a two-dimensional (2D) electron gas of high-mobility $\mathrm{(Al,Ga)As}$ heterostructures~\cite{Molenkamp-Jong:1994,Jong-Molenkamp:1995}. Among other effects, it was demonstrated that the resistivity decreases with temperature~\cite{Molenkamp-Jong:1994,Jong-Molenkamp:1995} if a hydrodynamic regime is realized. This phenomenon is known as the Gurzhi effect~\cite{Gurzhi:1963}. Later, the characteristic inverse-square dependence of the resistivity on the channel size was observed in the ultrapure 2D metal palladium cobaltate $\mathrm{PdCoO_2}$~\cite{Moll-Mackenzie:2016}, which agrees with the dependence expected for the Poiseuille flow of the electron fluid.

The realization of the hydrodynamic regime in graphene~\cite{Crossno-Fong:2016,Ghahari-Kim:2016,Krishna-Falkovich:2017,Berdyugin-Bandurin:2018,Bandurin-Falkovich:2018,Ku-Walsworth:2019,Sulpizio-Ilani:2019,Samaddar-Morgenstern:2021,Kumar-Ilani:2021} had a strong impact on the development of electron hydrodynamics in solids~\cite{Lucas-Fong:rev-2017,Narozhny:rev-2019}. Recently, signatures of the electron hydrodynamics were also observed in three-dimensional (3D) Dirac and Weyl semimetals. The experimental observation of the dependence of the electric resistivity on the channel width and the violation of the Wiedemann-Franz law with the lowest Lorentz number ever reported indicate the hydrodynamic transport regime in the Weyl semimetal WP$_2$~\cite{Gooth-Felser:2018}. Later, the hydrodynamic profile of electric current was visualized via stray magnetic fields in the Weyl semimetal WTe$_2$~\cite{Vool-Yacoby:2020-WTe2}. We notice that these observations, however, do not directly rely on the relativistic-like spectrum of Weyl semimetals and demonstrate universal properties of electron hydrodynamics. In addition to Dirac and Weyl semimetals, strongly correlated systems such as kagome metals may provide a platform for electron hydrodynamics~\cite{Sante-Wehling:2020}.

Motivated by the recent studies of electron hydrodynamics, we propose to probe the salient feature of Weyl and Dirac semimetals, namely, the chiral anomaly activated by an external static magnetic field, in crossover and hydrodynamic regimes of transport. While the chiral anomaly and the crossover regimes of transport were considered before, the corresponding studies were performed separately. In particular, the role of the chiral anomaly in a hydrodynamic regime was investigated in Ref.~\cite{Lucas-Sachdev:2016} in a high-energy physics approach; see also Refs.~\cite{Son-Surowka:2009,Isachenkov-Sadofyev:2011,Neiman-Oz:2011,Hidaka-Pu:2018} for the hydrodynamic framework for chiral plasmas. The finite-size effects, however, were not addressed. The transport properties of 2D systems with relativistic-like dispersion relation were studied in Ref.~\cite{Kashuba-Molenkamp:2018} in hydrodynamic and diffusive regimes in the absence of magnetic fields.

In this work, we combine two active research fields, namely the chiral anomaly and hydrodynamic transport. We predict that the part of the electric conductivity determined by the chiral anomaly (anomalous conductivity) exhibits an analog of the Gurzhi effect dubbed the \emph{anomalous Gurzhi effect}. In this case, the conductivity shows a nonmonotonic behavior with temperature: it decreases for small temperatures and grows for large ones. The anomalous Gurzhi effect has a similar origin to its conventional counterpart: in the hydrodynamic regime, the valley-imbalance (chiral) charge density, which quantifies the effect of the chiral anomaly, relaxes primarily at the boundaries rather than in the bulk. Hence, the anomalous conductivity is determined by the electron-electron collisions and the size of the system. Since the rate of the bulk internode scattering processes that relax the chiral charge is usually much smaller than that for the intranode ones, see, e.g., Refs.~\cite{Zhang-Hasan-TaAs:2016,Jadidi-Drew-TaAs:2019}, we expect that the anomalous Gurzhi effect should be easier to realize than the standard Gurzhi effect. As for the optimal experimental conditions, the anomalous Gurzhi effect might be observable in a clean film of a Weyl or Dirac semimetal subject to a magnetic field applied along the surface of the film. The surface of the film should be diffusive to allow for momentum and chirality relaxation.

It is worth noting that the mechanism of the anomalous Gurzhi effect resembles that of the anomalous nonlocal transport in Dirac and Weyl semimetals~\cite{Parameswaran-Vishwanath:2014,Zhang-Xiu:2017,Boer-Brinkman:2019}, where a valley-imbalance charge density induced by the chiral anomaly is transformed into an electric current away from the source. However, the role of boundaries and the emergence of a hydrodynamic regime were not investigated in these studies. An analogous mechanism is also responsible for the anomalous penetration and transmission of electromagnetic waves recently studied in Refs.~\cite{Sukhachov-Glazman:2021-skin,Matus-Surowka:2021}. Similar to the anomalous Gurzhi effect, the length scale for the nonlocal electromagnetic response is determined by the diffusion length of the valley charge imbalance. Furthermore, boundary conditions can strongly modify the valley-imbalance charge density near surfaces affecting a transmitted electromagnetic field~\cite{Sukhachov-Glazman:2021-skin}. The dependence of the transport properties on temperature, however, was not investigated in these works.

The paper is organized as follows. The phenomenology of the anomalous Gurzhi effect is provided in Sec.~\ref{sec:Simple}. In Sec.~\ref{sec:CKT}, we define the model and briefly describe the key equations of the chiral kinetic theory. The role of the chiral anomaly in the hydrodynamic regime is considered in Sec.~\ref{sec:Hydro}. Section~\ref{sec:Kinetic} is devoted to the study of transport properties in the kinetic approach. The results are summarized and discussed in Sec.~\ref{sec:Summary}. Technical details are given in several appendices. Throughout this study, we set $k_{B}=1$.

\section{Phenomenology of the anomalous Gurzhi effect}
\label{sec:Simple}

\begin{figure}[t]
\centering
\includegraphics[width=0.4\textwidth]{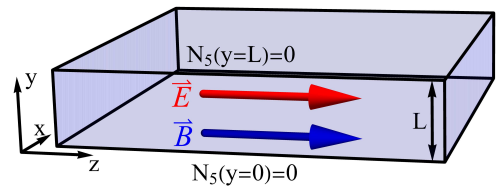}
\caption{
A schematic model setup where static electric $\mathbf{E}$ and magnetic $\mathbf{B}$ fields are applied along the film of a Weyl or Dirac semimetal. The thickness of the film along the $y$ direction is $L$ and the film is infinite along the $x$ and $z$ directions. We force the chiral charge density $N_5(y)$ to vanish at the surfaces of the film by the boundary conditions $N_5(y=0,L)=0$.
}
\label{fig:model}
\end{figure}

Let us start with a qualitative derivation of the Gurzhi effect and the anomalous Gurzhi effect. For illustrative purposes, we consider a model of a Weyl semimetal with two symmetric Weyl nodes. This is the case in, e.g., the Weyl semimetal EuCd$_2$As$_2$~\cite{Wang-Canfield:2019,Soh-Boothroyd:2019,Ma-Shi:2019}. The model setup is presented in Fig.~\ref{fig:model} where static electric  $\mathbf{E}$ and magnetic $\mathbf{B}$ fields are applied along the film of the semimetal. The thickness of the film along the $y$ direction is $L$ and the film is assumed to be infinite along the $x$ and $z$ directions. We make this assumption for simplicity. It should not matter much (in comparison to experiments) as long as the thickness of the film is much smaller than its width and length.

To study the Gurzhi effect and the anomalous Gurzhi effect, we calculate the electric current density $\mathbf{J}(t,\mathbf{r})$. The dynamics of this current is nontrivial in Weyl semimetals subject to a magnetic field and depends on valley-even (electric) and valley-odd (chiral) perturbations. Indeed, since there are two well-separated Weyl nodes (valleys), we can introduce valley-even and valley-odd charge ($N(t,\mathbf{r})$ and $N_5(t,\mathbf{r})$, respectively) and current ($\mathbf{J}(t,\mathbf{r})$ and $\mathbf{J}_5(t,\mathbf{r})$, respectively) densities. The corresponding continuity equations read
\begin{eqnarray}
\label{Simple-N}
&&\partial_t N(t,\mathbf{r}) +\left(\bm{\nabla}\cdot\mathbf{J}(t,\mathbf{r})\right)\!=\! 0,\\
\label{Simple-N5}
&&\partial_t N_5(t,\mathbf{r}) +\left(\bm{\nabla}\cdot\mathbf{J}_5(t,\mathbf{r})\right)\!= \! -\frac{e^3\left(\mathbf{E}\cdot\mathbf{B}\right) }{2\pi^2 \hbar^2 c} - \frac{N_5(t,\mathbf{r})}{\tau_{\rm ei, 5}},
\end{eqnarray}
see also Refs.~\cite{Burkov:rev-2015,Parameswaran-Vishwanath:2014}. The electric and chiral currents are
\begin{eqnarray}
\label{Simple-J-def}
\mathbf{J}(t,\mathbf{r}) &=& \sigma_0 \mathbf{E} -D_{\rm eff}\bm{\nabla} N(t,\mathbf{r}) +\frac{e^2 \mu_5(t,\mathbf{r})}{2\pi^2 \hbar^2c} \mathbf{B},\\
\label{Simple-J5-def}
\mathbf{J}_5(t,\mathbf{r}) &=& -D_{\rm eff}\bm{\nabla} N_5(t,\mathbf{r}) +\frac{e^2 \mu(t,\mathbf{r})}{2\pi^2\hbar^2c} \mathbf{B}.
\end{eqnarray}
Here, $\sigma_0 = 2e^2 \nu(\mu) D_{\rm eff}$ is the electric conductivity, $\nu(\mu)$ is the density of states (DOS) per Weyl node, and $D_{\rm eff}=v_F^2\tau_{\rm eff}/3$ is the diffusion constant determined by the effective scattering time $\tau_{\rm eff}$, whose meaning will be clarified later. Notice that in the model with symmetric Weyl nodes, the parameters such as the Fermi velocity $v_F$, the DOS, and the effective scattering time are the same in all nodes. In addition, $\mu(t,\mathbf{r})$ and $\mu_5(t,\mathbf{r})$ are electric and chiral chemical potentials that quantify the corresponding charge densities. The continuity relations and the corresponding currents will be discussed in more detail in Sec.~\ref{sec:Hydro-eqs}.

The currents (\ref{Simple-J-def}) and (\ref{Simple-J5-def}) have a standard form for the diffusive regime albeit contain contributions related to the chiral anomaly known as chiral magnetic and chiral separation effect currents~\cite{Vilenkin:1980,Metlitski:2005,Fukushima:2008}; see the terms $\propto \mathbf{B}$. These currents are crucial to couple dynamics of electric and chiral charges. Notice also that, in addition to the bulk internode electron-impurity scattering quantified by $\tau_{\rm ei, 5}$, the chiral charge density $N_5(t,\mathbf{r})$ is not conserved due to the anomalous $\propto \left(\mathbf{E}\cdot\mathbf{B}\right)$ contribution in Eq.~(\ref{Simple-N5}); see, e.g., Ref.~\cite{Nielsen:1983}. This is a new element brought by the topologically nontrivial electronic band structure of Weyl and certain Dirac semimetals.

Let us first consider the conventional Gurzhi effect at $B=0$. Notice that, in the film geometry, see Fig.~\ref{fig:model}, all variables depend only on $y$ (i.e., the coordinate along the surface normal in the film). It is easy to verify that the direct electric current along the surface of the film reads as $J_{\parallel}=\sigma_0 E_{\parallel} \sim \tau_{\rm eff} E_{\parallel}$; see Eq.~(\ref{Simple-J-def}). If the intranode scattering is strong, i.e., the intranode electron-impurity scattering length $l_{\rm ei} =v_F \tau_{\rm ei}$ is much smaller than the film width $L$, the internode scattering length $l_{\rm ei,5}=v_F \tau_{\rm ei,5}$, and the electron-electron scattering length $l_{\rm ee}=v_F \tau_{\rm ee}$, we have a conventional Drude relation for the conductivity with $\tau_{\rm eff}=\tau_{\rm ei}$. The conductivity depends weakly on temperature in this case.

However, as was conceived by Gurzhi in Ref.~\cite{Gurzhi:1963}, the hydrodynamic-like transport regime with $\tau_{\rm eff}=L^2/\tau_{\rm ee}$ is possible if $l_{\rm ee}\ll L, l_{\rm ei}, l_{\rm ei,5}$ and the scattering at the surfaces of the film is diffusive. Therefore, since $l_{\rm ee}\propto 1/T^2$ for $T\ll\mu$, the electric conductivity $\sigma_0 \propto 1/l_{\rm ee}$ rises with temperature. This phenomenon is known as the conventional Gurzhi effect.

As Gurzhi pointed out, the same result can be obtained in the hydrodynamic approach by introducing the drift velocity $\mathbf{u}(y)$ and solving the Navier-Stokes equation for a steady flow:
\begin{equation}
\label{Simple-NS}
\eta_{\rm kin} \partial_y^2 \mathbf{u}(y)- \frac{\mathbf{u}(y)}{\tau_{\rm ei}} = \frac{e}{m_{\rm eff}}\mathbf{E}.
\end{equation}
Here, $\eta_{\rm kin} \propto \tau_{\rm ee}$ is the kinematic viscosity, $\mathbf{u}(y)/\tau_{\rm ei}$ describes the momentum relaxation of the fluid, and $m_{\rm eff}$ is the effective mass whose explicit expression is not important now. The solution to Eq.~(\ref{Simple-NS}) reads
\begin{equation}
\label{Simple-NS-sol}
u_{\parallel}(y)= -\frac{e E \tau_{\rm ei}}{m_{\rm eff}}\left[1 - \frac{\cosh{\left(\frac{L-2y}{2\lambda_{\rm G}}\right)}}{\cosh{\left(\frac{L}{2\lambda_{\rm G}}\right)}}\right],
\end{equation}
where we introduce the Gurzhi length $\lambda_{\rm G} = \sqrt{\eta_{\rm kin} \tau_{\rm ei}}$ and assume no-slip boundary conditions $u_{\parallel}(y=0,L)=0$. Then, by using the hydrodynamic expression for the electric current $\mathbf{J}(y)=N\mathbf{u}(y)$ and averaging over the film width, we obtain the following averaged conductivity:
\begin{equation}
\label{Simple-NS-sigma-sol}
\sigma_0 \propto
\begin{cases}
    \frac{L^2}{12\eta_{\rm kin}}, & \lambda_{\rm G} \gg L,\\
    \tau_{\rm ei}, & \lambda_{\rm G} \ll L.
  \end{cases}
\end{equation}
The case $\lambda_{\rm G} \gg L$ corresponds to the hydrodynamic regime with $\sigma_0 \propto 1/l_{\rm ee}$.

\begin{figure}[t]
\centering
\includegraphics[width=0.4\textwidth]{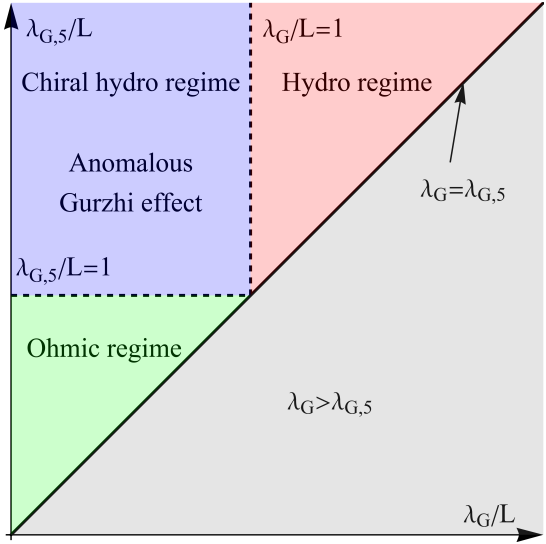}
\caption{
Transport phase diagram of the system with three main regimes: (i) Ohmic ($\lambda_{\rm G}\ll L$ and $\lambda_{\rm G, 5}\ll L$), (ii) chiral hydrodynamic ($\lambda_{\rm G}\ll L$ and $\lambda_{\rm G, 5}\gg L$), and (iii) hydrodynamic ($\lambda_{\rm G}\gg L$ and $\lambda_{\rm G, 5}\gg L$). Here, $\lambda_{\rm G}\sim \sqrt{l_{\rm ee} l_{\rm ei}}$ and $\lambda_{\rm G,5}\sim \sqrt{l_{\rm ee} l_{\rm ei,5}}$ are Gurzhi and chiral Gurzhi lengths, respectively, where $l_{\rm ei}$, $l_{\rm ei,5}$, and $l_{\rm ee}$ are the intranode, internode, and electron-electron scattering lengths, respectively. Since the intranode electron-impurity scattering length is usually smaller than the internode one~\cite{Zhang-Hasan-TaAs:2016,Jadidi-Drew-TaAs:2019}, i.e., $l_{\rm ei} \ll l_{\rm ei,5}$, the regime $\lambda_{\rm G}\gtrsim \lambda_{\rm G, 5}$ (gray shaded region) cannot be realized.
}
\label{fig:MP-cor-phase-diag}
\end{figure}

Now, let us turn our attention to the case with a nonvanishing magnetic field $B\neq0$ and focus on the anomalous part of the electric current given by the last term in Eq.~(\ref{Simple-J-def}). This part of the current is quantified by the valley-imbalance charge density $N_{5}(y)$. To find the latter, we rewrite Eq.~(\ref{Simple-N5}) as
\begin{equation}
\label{Simple-N5-1}
D_{\rm eff}\partial_y^2 N_{5}(y) -\frac{N_5(y)}{\tau_{\rm ei, 5}} =  \frac{e^3 \left(\mathbf{E}\cdot\mathbf{B}\right)}{2\pi^2 \hbar^2c}.
\end{equation}
Its solution for the chirality-mixing boundary conditions $N_5(y=0,L)=0$ resembles that for the hydrodynamic velocity (\ref{Simple-NS-sol}), i.e.,
\begin{equation}
\label{Simple-N5-1-sol}
N_5(y)= -\frac{e^3 \left(\mathbf{E}\cdot\mathbf{B}\right) \tau_{\rm ei, 5}}{2\pi^2 \hbar^2c} \left[1 - \frac{\cosh{\left(\frac{L-2y}{2\lambda_{\rm G,5}}\right)}}{\cosh{\left(\frac{L}{2\lambda_{\rm G,5}}\right)}}\right],
\end{equation}
where we introduced the chiral Gurzhi length $\lambda_{\rm G,5} = \sqrt{D_{\rm eff} \tau_{\rm ei, 5}}$. This result allows us to find the anomalous part of the conductivity averaged over the film width:
\begin{equation}
\label{Simple-N5-1-sigma-sol}
\sigma_{\rm anom} \propto B^2
  \begin{cases}
    \frac{L^2}{12D_{\rm eff}}, & \lambda_{\rm G,5} \gg L,\\
    \tau_{\rm ei,5}, & \lambda_{\rm G,5} \ll L.
  \end{cases}
\end{equation}
Evidently, the behavior of the anomalous part of the conductivity resembles that of the normal one given in Eq.~(\ref{Simple-NS-sigma-sol}) albeit is controlled by different length scales. Since $\tau_{\rm ei,5}\gg\tau_{\rm ei}$, it is possible to realize the regime $\lambda_{\rm G} \ll L$ and $\lambda_{\rm G,5} \gg L$ where the Gurzhi effect appears only in the anomalous part of the conductivity. Therefore, we dub this phenomenon the anomalous Gurzhi effect. We summarize the main transport regimes in Fig.~\ref{fig:MP-cor-phase-diag}.

The crucial difference between the anomalous $\sigma_{\rm anom}$ and normal $\sigma_0$ parts of the conductivity is that while the latter is determined by momentum-relaxation processes, the former relies on the chiral charge relaxation (or valley imbalance equilibration). We consider strong ($l_{\rm ei,5}\ll L$) and weak ($l_{\rm ei,5}\gg  L$) relaxation of the chiral charge in the bulk. In the former case, we have the conventional result $\sigma_{\rm anom} \propto \tau_{\rm ei, 5}$~\cite{Son-Spivak:2013}. However, when the bulk internode scattering is weak ($l_{\rm ei,5}\gg  L$), the chirality relaxation occurs primarily at the surfaces of the film due to the boundary conditions $N_5(y=0,L)=0$. In order to reach the surface, electrons move diffusively across the film with the diffusion coefficient determined either by intranode or electron-electron scattering rates. Therefore, the qualitative result in Eq.~(\ref{Simple-N5-1-sigma-sol}) is applicable in both dirty and clean regimes. The anomalous Gurzhi effect appears only in the clean case.

Thus, the anomalous Gurzhi effect originates from the diffusion of the chiral charge density due to the electron-electron collisions and the relaxation of the chiral charge that occurs predominantly at the boundaries.

\section{Chiral kinetic theory, boundary conditions, and collision integrals}
\label{sec:CKT}

\subsection{Kinetic equations and boundary conditions}
\label{sec:CKT-eqs}

In this section, we present the key equations of the chiral kinetic theory (CKT) that will be used in the study of the transport properties of the film. The current density per Weyl node $\alpha$ is defined as~\cite{Xiao-Niu:rev-2010,Son:2013,Stephanov:2012,Son-Spivak:2013}
\begin{eqnarray}
\label{CKT-cc-j}
\mathbf{J}_{\alpha}(t,\mathbf{r}) &=& -e\sum_{\eta}\eta \int \frac{d^3p}{(2\pi \hbar)^3} \Big\{\mathbf{v}_{\alpha,\eta} -e\left[\tilde{\mathbf{E}}_{\alpha,\eta}\times\mathbf{\Omega}_{\alpha,\eta}\right] \nonumber\\
&-&\frac{e}{c}(\mathbf{v}_{\alpha,\eta}\cdot\mathbf{\Omega}_{\alpha,\eta})\mathbf{B}\Big\} f_{\alpha,\eta}(t,\mathbf{r},\mathbf{p}) \nonumber\\
&\approx&
e \sum_{\eta}\eta \int \frac{d^3p}{(2\pi \hbar)^3} \Big[\mathbf{v}_{\alpha,\eta} -\frac{e}{c}(\mathbf{v}_{\alpha,\eta}\cdot\mathbf{\Omega}_{\alpha,\eta})\mathbf{B}\Big] \nonumber\\
&\times& n_{\alpha,\eta}(t,\mathbf{r},\mathbf{p}) f_{\alpha,\eta}^{\prime}(\mathbf{p}).
\end{eqnarray}
In the last equation, we left only the terms linear in the perturbed distribution function~\footnote{The term $\propto e\left[\tilde{\mathbf{E}}_{\alpha,\eta}\times\mathbf{\Omega}_{\alpha,\eta}\right]f_{\alpha,\eta}^{(0)}(\mathbf{p})$ in Eq.~(\ref{CKT-cc-j}) corresponds to the anomalous Hall effect~\cite{Yang-Lu:2011,Burkov-Balents:2011}. It is irrelevant for our study of the chiral anomaly in time-reversal symmetric materials or for certain orientations of the Weyl nodes in materials with broken time-reversal symmetry. Some of the effects related to the anomalous Hall effect in hydrodynamic regime in Weyl semimetals are discussed in Refs.~\cite{Gorbar:2018vuh,Gorbar:2018sri}.}
\begin{equation}
\label{CKT-equation-f-chi}
f_{\alpha,\eta}(t,\mathbf{r};\mathbf{p}) \approx  f_{\alpha,\eta}^{(0)}(\mathbf{p})- n_{\alpha,\eta}(t,\mathbf{r};\mathbf{p}) f_{\alpha,\eta}^{\prime}(\mathbf{p}),
\end{equation}
where $f_{\alpha,\eta}^{(0)}(\mathbf{p})$ is the Fermi-Dirac distribution function, $f_{\alpha,\eta}^{\prime}(\mathbf{p})$ is its derivative with respect to energy, and $\eta=\pm$ corresponds to conduction ($\eta=+$) and valence ($\eta=-$) bands.
The salient property of the CKT is the presence of Berry curvature $\mathbf{\Omega}_{\alpha,\eta}=\mathbf{\Omega}_{\alpha,\eta}(\mathbf{p})$ that quantifies the nontrivial topological properties in the semiclassical approach. Further, $\mathbf{v}_{\alpha,\eta}=\partial_{\mathbf{p}}\tilde{\epsilon}_{\alpha,\eta}$ is the quasiparticle velocity, $\tilde{\epsilon}_{\alpha,\eta}=\tilde{\epsilon}_{\alpha,\eta}(\mathbf{p})$ is the quasiparticle dispersion relation at the node $\alpha$, $\mathbf{B}$ is the external magnetic field, and $\tilde{\mathbf{E}}_{\alpha,\eta} = \mathbf{E} +(1/e)\bm{\nabla}\tilde{\epsilon}_{\alpha,\eta}$ is the effective electric field, which includes the contribution from the band dispersion and the screened electric field. There is an additional Zeeman-like term in the quasiparticle energy dispersion relation $\tilde{\epsilon}_{\alpha,\eta} = \epsilon_{\alpha,\eta}-\left(\mathbf{B}\cdot\mathbf{m}_{\alpha,\eta}\right)$~\cite{Chang-Niu:1996,Son:2013,Xiao-Niu:rev-2010}, where $\mathbf{m}_{\alpha,\eta}$ is the orbital magnetic moment~\footnote{In addition to the magnetic moment correction $\left(\mathbf{B}\cdot\mathbf{m}_{\alpha,\eta}\right)$, there are also terms quadratic in magnetic field that originate from the inter-band corrections~\cite{Gao-Niu:2015,Gorbar:2017cwv}.}. The simultaneous presence of the Berry curvature and magnetic field modifies the phase-space volume~\cite{Xiao-Niu:rev-2010}, which is quantified by the term $\Theta_{\alpha,\eta}=\left[1-e \left(\mathbf{B}\cdot \mathbf{\Omega}_{\alpha,\eta}\right)/c\right]$. In addition, $-e$ is the charge of the electron and $c$ is the speed of light.

Since we are interested in the effects of the chiral anomaly on the transport properties, the magnetic moment $\mathbf{m}_{\alpha,\eta}$ and the renormalization of the phase-space volume $\Theta_{\alpha,\eta}$ can be neglected. As we discuss in Appendix~\ref{sec:app-MP-cor}, these terms produce only subleading corrections for certain transport regimes. This approximation significantly simplifies the calculations. By using the expansion (\ref{CKT-equation-f-chi}), the Boltzmann equation of the CKT reads~\cite{Xiao-Niu:rev-2010,Son:2013,Stephanov:2012,Son-Spivak:2013}
\begin{widetext}
\begin{eqnarray}
\label{CKT-equation-kinetic-equation-simpl}
&&\partial_t n_{\alpha,\eta}(t,\mathbf{r};\mathbf{p}) f_{\alpha,\eta}^{\prime}(\mathbf{p})
+\left[\mathbf{v}_{\alpha,\eta} -\frac{e}{c}(\mathbf{v}_{\alpha,\eta}\cdot\mathbf{\Omega}_{\alpha,\eta})\mathbf{B}\right]\cdot \bm{\nabla} n_{\alpha,\eta}(t,\mathbf{r};\mathbf{p}) f_{\alpha,\eta}^{\prime}(\mathbf{p})
-\frac{e}{c}\left[\mathbf{v}_{\alpha,\eta}\times \mathbf{B}\right] \cdot \left(\partial_{\mathbf{p}} n_{\alpha,\eta}(t,\mathbf{r};\mathbf{p})\right) f_{\alpha,\eta}^{\prime}(\mathbf{p}) \nonumber\\
&&=-e \left[\left(\mathbf{v}_{\alpha,\eta}\cdot\mathbf{E}\right) -\frac{e}{c}\left(\mathbf{E}\cdot\mathbf{B}\right)\left(\mathbf{v}_{\alpha,\eta}\cdot\mathbf{\Omega}_{\alpha,\eta}\right)\right]f_{\alpha,\eta}^{\prime}(\mathbf{p})
-I\left[n_{\alpha,\eta}\right],
\end{eqnarray}
\end{widetext}
where the collision integral is denoted by $I\left[n_{\alpha,\eta}\right]$.

In the case of the film geometry shown in Fig.~\ref{fig:model}, we assume diffusive boundary conditions where quasiparticles from different Weyl nodes intermix at the surfaces $y=0,L$; i.e., the internode relaxation is fast at the boundaries. These boundary conditions are given by
\begin{eqnarray}
\label{CKT-BC-y=0}
v_{\alpha,\eta,y}>0:\,\,  &&n_{\alpha,\eta}(t,\mathbf{r}; \mathbf{p})\Big|_{y=0} = \frac{1}{2\pi N_{W}} \sum_{\beta}^{N_{W}}\int_{0}^{2\pi}\! d\varphi^{\prime} \int_{0}^{\pi}\! d\theta^{\prime}\nonumber\\
&&\times\sin{(\theta^{\prime})}\, \theta\left[-v_{\beta,\eta,y}(p, \theta^{\prime}, \varphi^{\prime})\right] \nonumber\\
&&\times n_{\beta,\eta}(t,\mathbf{r}; p, \theta^{\prime}, \varphi^{\prime})\Big|_{y=0}
\end{eqnarray}
and
\begin{eqnarray}
\label{CKT-BC-y=L}
v_{\alpha,\eta,y}<0:\,\,  &&n_{\alpha,\eta}(t,\mathbf{r}; \mathbf{p})\Big|_{y=L} = \frac{1}{2\pi N_{W}}\sum_{\beta}^{N_{W}}\int_{0}^{2\pi}\! d\varphi^{\prime} \int_{0}^{\pi}\! d\theta^{\prime}\nonumber\\
&&\times\sin{(\theta^{\prime})}\, \theta\left[v_{\beta,\eta,y}(p, \theta^{\prime}, \varphi^{\prime})\right] \nonumber\\
&&\times n_{\beta,\eta}(t,\mathbf{r}; p, \theta^{\prime}, \varphi^{\prime})\Big|_{y=L};
\end{eqnarray}
see also Refs.~\cite{Beenakker-Houten:1991,Jong-Molenkamp:1995,Kashuba-Molenkamp:2018} for the boundary conditions in two-dimensional systems. Here, $v_{\alpha,\eta,y}$ is the $y$ component of the quasiparticle velocity (i.e., normal to the surface of the film), $N_{W}$ is the number of Weyl nodes, and $\theta(x)$ is the Heaviside function. The boundary conditions (\ref{CKT-BC-y=0}) and (\ref{CKT-BC-y=L}) mean that quasiparticles moving away from the boundaries have isotropic angular distribution. Also, the numbers of impinging and outgoing quasiparticles are the same. By multiplying Eqs.~(\ref{CKT-BC-y=0}) and (\ref{CKT-BC-y=L}) with the chirality of a Weyl node and summing over all nodes, it is straightforward to see that the phenomenological chirality-mixing boundary conditions $N_{5}(y=0,L)=0$ introduced in Sec.~\ref{sec:Simple} agree with those in Eqs.~(\ref{CKT-BC-y=0}) and (\ref{CKT-BC-y=L}).

\subsection{Collision integral}
\label{sec:CKT-coll}

Let us discuss the key features of the collision integral $I\left[n_{\alpha,\eta}\right]$. The details are presented in Appendix~\ref{sec:app-I-coll}. We take into account the scattering on disorder with local potential and electron-electron collisions. In the relaxation time approximation, the part of the collision integral responsible for the electron-impurity scattering reads
\begin{equation}
\label{CKT-coll-I-coll}
I_{\rm ei}\left[n_{\alpha,\eta}(t,\mathbf{r};\mathbf{p})\right] \approx \sum_{\beta}^{N_{W}}\frac{n_{\alpha,\eta}(t,\mathbf{r};\mathbf{p}) -\overline{n_{\beta,\eta}}(t,\mathbf{r};p)}{\tau_{\alpha,\beta}(p)} f_{\alpha,\eta}^{\prime}(\mathbf{p}),
\end{equation}
where $1/\tau_{\alpha,\beta}(p)$ is the electron-impurity scattering rate and $\overline{n_{\beta,\eta}}(t,\mathbf{r};p)$ is the distribution function averaged over the momentum direction~\footnote{In Eq.~(\ref{CKT-coll-I-coll}), we neglect the angular dependence in the matrix element of the disorder potential. This does not qualitatively change our results.}. Intra- and internode scatterings correspond to terms with $\alpha=\beta$ and $\alpha\neq\beta$ in Eq.~(\ref{CKT-coll-I-coll}), respectively.

In general, due to the presence of several nodes, the scattering integral couples all $N_{W}$ kinetic equations. To simplify our presentation, we consider a Weyl semimetal in the presence of a symmetry between the pairs ($\alpha,-\alpha$) of Weyl nodes of opposite chiralities. Moreover, we assume that the pairs are well separated and that the electron dispersion around each of the nodes is determined by the same parameters. In this case, $\tau_{\alpha,\alpha}(p)=\tau_{\rm ei}(p)$ and $\tau_{\alpha,-\alpha}(p)=2\tau_{\rm ei, 5}(p)$. The model with two Weyl nodes used in Sec.~\ref{sec:Simple} is the simplest example of such a Weyl semimetal.

Let us now turn our attention to electron-electron scattering. We use the Callaway ansatz~\cite{Callaway:1959,Jong-Molenkamp:1995}:
\begin{eqnarray}
\label{CKT-coll-I-ee-fin}
I_{\rm ee}\left[n_{\alpha,\eta}(t,\mathbf{r};\mathbf{p})\right] &=& \frac{n_{\alpha,\eta}(t,\mathbf{r};\mathbf{p}) -\left\langle n_{\alpha}(t,\mathbf{r})\right\rangle -\left(\mathbf{p}\cdot\mathbf{u}(t,\mathbf{r})\right)}{\tau_{\rm ee}} \nonumber\\ &\times&f_{\alpha,\eta}^{\prime}(\mathbf{p}),
\end{eqnarray}
where
\begin{eqnarray}
\label{CKT-coll-I-ee-aver-def}
\left\langle n_{\alpha}(t,\mathbf{r})\right\rangle &=&  \frac{1}{\sum_{\eta}\eta
\int d^3p\, f_{\alpha,\eta}^{\prime}(p)} \nonumber\\
&\times&\sum_{\eta}\eta \int d^3p\, n_{\alpha,\eta}(t,\mathbf{r};\mathbf{p}) f_{\alpha,\eta}^{\prime}(\mathbf{p}),\\
\label{CKT-coll-I-ee-u}
\mathbf{u}(t,\mathbf{r}) &=& \frac{3}{\sum_{\eta} \eta \sum_{\alpha}^{N_{W}} \int d^3p\, p^2f_{\alpha,\eta}^{\prime}(p)} \nonumber\\
&\times&\sum_{\eta}\! \eta \sum_{\alpha}^{N_{W}}\! \int d^3p\, \mathbf{p}\, n_{\alpha,\eta}(t,\mathbf{r};\mathbf{p}) f_{\alpha,\eta}^{\prime}(\mathbf{p}).
\end{eqnarray}
Notice that the Callaway ansatz is phenomenological and does not follow from microscopic models. It has limitations of the relaxation time approximation such as the independence of the relaxation time on distribution functions. However, it captures the main properties of electron-electron scattering, namely the conservation of  electric charge per Weyl node and total momentum. Furthermore, it significantly simplifies the solution of the kinetic equations.

\subsection{Linearized model}
\label{sec:CKT-lin}

While the kinetic approach discussed before is general and does not rely on the relativistic-like linear energy spectrum of Weyl and Dirac semimetals, in our explicit calculations and numerical estimates, we use a minimal model with an isotropic linear spectrum. The effective low-energy Weyl Hamiltonian in the vicinity of the Weyl node $\alpha$ reads as
\begin{equation}
\label{CKT-lin-H-def}
H_{\alpha} =\chi_{\alpha} v_{F} \left(\mathbf{p}\cdot\bm{\sigma}\right),
\end{equation}
where $\chi_{\alpha} =\pm$ is the chirality or, equivalently, the topological charge of the Weyl node, $v_{F}$ is the Fermi velocity, and $\bm{\sigma}$ is the vector of the Pauli matrices acting on pseudospin space. In this model, the Berry curvature $\mathbf{\Omega}_{\alpha,\eta}$ takes the form
\begin{equation}
\label{CKT-lin-Berry-monopole}
\mathbf{\Omega}_{\alpha, \eta} =\eta \chi_{\alpha} \hbar\frac{\mathbf{p}}{2p^3}
\end{equation}
and the DOS per Weyl node $\nu_{\alpha}(p)=\nu(p)$ is
\begin{equation}
\label{CKT-lin-DOS}
\nu(p) = \frac{p^2}{2\pi^2 \hbar^3v_{F}}.
\end{equation}
In addition, since the quasiparticle energy does not depend on the nodal index and the momentum direction, $\epsilon_{\alpha,\eta}=\eta v_F p$, we have $f_{\alpha,\eta}^{(0)}(\mathbf{p})=f_{\eta}^{(0)}(p)$, $f_{\alpha,\eta}^{\prime}(p)=f_{\eta}^{\prime}(p)$, and $\mathbf{v}_{\alpha}=\mathbf{v}$.

Finally, let us discuss the explicit expressions for the relaxation times in the linear model. For the local electron-impurity scattering potential, the intranode scattering time reads as
\begin{equation}
\label{CKT-lin-tau-ei}
\tau_{\rm ei}(p) = \tau_{\rm ei}(p_F)\frac{p_F^2}{p^2},
\end{equation}
where $p_F=\mu/v_F$ is the Fermi momentum. The same expression albeit with $\tau_{\rm ei}(p_F)\to \tau_{\rm ei,5}(p_F)$ holds for the internode scattering time $\tau_{\rm ei,5}(p)$. The intranode scattering time averaged over the Fermi surface is given by
\begin{equation}
\label{CKT-lin-tau-ei-aver}
\tau_{\rm ei} = \left\langle \tau_{\rm ei}(p) \right\rangle= \tau_{\rm ei}(p_F) \frac{\mu^2}{\mu^2+\pi^2T^2/3};
\end{equation}
see Eq.~(\ref{CKT-coll-I-ee-aver-def}) and Appendix~\ref{sec:app-1}. The temperature dependence of $\tau_{\rm ei}$ has no qualitative effect on the conventional and anomalous Gurzhi effects for $T\lesssim \mu$ and may be omitted. However, in order to obtain the temperature-independent resistivity in the Ohmic transport regime, we retain the dependence on $T$ in Eq.~(\ref{CKT-lin-tau-ei-aver}).

As for the electron-electron scattering time, we use the following standard Fermi liquid expression:
\begin{equation}
\label{CKT-lin-tau-ee-def}
\tau_{\rm ee} = \frac{\hbar}{\mu \alpha_{\rm eff}^2} \frac{\mu^2}{T^2},
\end{equation}
which is qualitatively valid for $T\lesssim \mu$. Here, $\alpha_{\rm eff}$ is the effective fine-structure constant that includes the effects of screening. Unlike the electron-impurity scattering time, the electron-electron scattering time changes rapidly with $T$. In what follows, to simplify the calculations, we employ scattering times averaged over the Fermi surface.

\section{Hydrodynamic approach}
\label{sec:Hydro}

\subsection{General equations}
\label{sec:Hydro-eqs}

To study the transport properties of the film of Dirac or Weyl semimetal in a magnetic field, we start with the hydrodynamic approach. The hydrodynamic equations correspond to the conservation of total momentum, electric charge density per Weyl node, and energy. The details of their derivation for the case $\tau_{\rm ee}\ll \tau_{\rm ei}, \tau_{\rm ei,5}$ are given in Appendix~\ref{sec:app-hydro}. In this section, we focus on the longitudinal response. Then, the Navier-Stokes equation for quasiparticles with a relativistic-like dispersion relation reads
\begin{equation}
\label{hydro-eqs-NS}
\frac{w_0}{v_F^2} \partial_t u_{\parallel}(t,\mathbf{r}) +\frac{w_0}{v_F^2 \tau_{\rm eff}} u_{\parallel}(t,\mathbf{r}) -\eta_{\rm dyn} \Delta u_{\parallel}(t,\mathbf{r}) = N_0 E.
\end{equation}
Here, $u_{\parallel}(t,\mathbf{r})= \left(\hat{\mathbf{E}}\cdot \mathbf{u}(t,\mathbf{r})\right)$, $\hat{\mathbf{E}}=\mathbf{E}/E$ is the unit vector in the direction of the electric field, $\eta_{\rm dyn}  = w_0\tau_{\rm ee}/5$ is the dynamic viscosity~\footnote{Since we focus on the case where electric and magnetic fields are directed along the film, the Hall-type components of the viscosity tensor, which can appear in a magnetic field, do not affect the transport properties and can be omitted.}, $w_0=4\epsilon_0/3$ is the enthalpy density, $\epsilon_0$ is the equilibrium energy density, and $1/\tau_{\rm eff}=1/\tau_{\rm ei}+1/(2\tau_{\rm ei,5})$ is the scattering rate corresponding to momentum relaxation.

We introduce the time- and coordinate-dependent valley-even (electric) $N(t,\mathbf{r})=\sum_{\alpha}^{N_{W}}N_{\alpha}(t,\mathbf{r})$ and valley-odd (chiral) $N_5(t,\mathbf{r})=\sum_{\alpha}^{N_{W}}\chi_{\alpha}N_{\alpha}(t,\mathbf{r})$ charge densities. They satisfy the continuity equations (\ref{Simple-N}) and (\ref{Simple-N5}), respectively. In the latter, however, the chiral anomaly term should be multiplied by $N_{W}/2$. In the global equilibrium state, $N(t,\mathbf{r})=N_0$ and $N_{5}(t,\mathbf{r})=0$. The explicit expressions for the equilibrium thermodynamic variables are given in Appendix~\ref{sec:app-1}.

The linearized electric $\mathbf{J}(t,\mathbf{r})$ and chiral $\mathbf{J}_5(t,\mathbf{r})$ current densities read
\begin{eqnarray}
\label{hydro-eqs-J}
\mathbf{J}(t,\mathbf{r}) &=& N_0 \mathbf{u}(t,\mathbf{r}) -D_{\rm ee}\bm{\nabla} N(t,\mathbf{r}) -\mathbf{v}_{\Omega}N_{5}(t,\mathbf{r}),\\
\label{hydro-eqs-J5}
\mathbf{J}_{5}(t,\mathbf{r}) &=& -D_{\rm ee}\bm{\nabla} N_{5}(t,\mathbf{r}) -\mathbf{v}_{\Omega}N(t,\mathbf{r}),
\end{eqnarray}
where we use $D_{\rm ee}=v_F^2 \tau_{\rm ee}/3$ and define the anomalous velocity $\mathbf{v}_{\Omega}$ as
\begin{equation}
\label{hydro-vOmega-def}
\mathbf{v}_{\Omega}= \frac{e}{4\pi^2 \hbar^2 c \nu_{\rm eff}(\mu)} \mathbf{B}.
\end{equation}
Here, the effective DOS $\nu_{\rm eff}(\mu)$ for the linearized spectrum reads
\begin{equation}
\label{Hydro-eqs-nu-eff-def}
\nu_{\rm eff}(\mu) = \frac{\mu^2+\pi^2T^2/3}{2\pi^2 \hbar^3 v_F^3}.
\end{equation}
Notice that the terms $\sim \tau_{\rm ee}$, e.g., the viscosity $\eta_{\rm dyn}$ in Eq.~(\ref{hydro-eqs-NS}) and diffusion coefficient $D_{\rm ee}$ in Eqs.~(\ref{hydro-eqs-J}) and (\ref{hydro-eqs-J5}), require us to calculate first-order corrections to the hydrodynamic equations; see Appendix~\ref{sec:app-hydro} for details.

In what follows, we consider the steady regime and the film geometry, see Fig.~\ref{fig:model}, where the hydrodynamic equations are rewritten as
\begin{eqnarray}
\label{hydro-eqs-NS-steady}
&&\frac{w_0}{v_F^2 \tau_{\rm eff}}u_{\parallel}(y) -\eta_{\rm dyn} \partial_y^2 u_{\parallel}(y) = N E,\\
\label{Hydro-eqs-N-steady}
&&D_{\rm ee}\partial_y^2 N(y) = 0,\\
\label{Hydro-eqs-N5-steady}
&&D_{\rm ee}\partial_y^2 N_{5}(y)-\frac{N_{5}(y)}{\tau_{\rm ei,5}} = -N_{W}e^2\nu_{\rm eff}(\mu) v_{\Omega} E.
\end{eqnarray}
These equations should be amended with boundary conditions. We employ no-slip boundary conditions (realized for rough boundaries) where the electron fluid sticks to the boundaries, $u_{\parallel}(y=0,L)=0$~\cite{Landau:t6-2013}. As for $N(y)$ and $N_5(y)$, we require that neither electric nor chiral charge is accumulated at the boundaries, $N(y=0,L)=0$ and $N_5(y=0,L)=0$. This agrees with the absence of surface charges and the boundary conditions (\ref{CKT-BC-y=0}) and (\ref{CKT-BC-y=L}).

\subsection{Solutions and conductivity}
\label{sec:Hydro-cond}

Let us now solve Eqs.~(\ref{hydro-eqs-NS-steady}), (\ref{Hydro-eqs-N-steady}), and (\ref{Hydro-eqs-N5-steady}). As one can see, these equations decouple and can be solved independently. We start with the Navier-Stokes equation (\ref{hydro-eqs-NS-steady}), whose solution is given by
\begin{equation}
\label{Hydro-cond-uz}
u_{\parallel}(y) = \frac{v_F^2 \tau_{\rm eff} N_0 E}{w_0} \left[1 - \frac{\cosh{\left(\frac{L-2y}{2\lambda_{\rm G}}\right)}}{\cosh{\left(\frac{L}{2\lambda_{\rm G}}\right)}}\right],
\end{equation}
where
\begin{equation}
\label{Hydro-cond-lambda-G}
\lambda_{\rm G} = \sqrt{v_F^2 \frac{\eta_{\rm dyn} \tau_{\rm eff}}{w_0}} =\sqrt{\frac{l_{\rm ee} l_{\rm eff}}{5}}
\end{equation}
is the Gurzhi length. The scattering lengths are defined by multiplying the corresponding scattering times by the Fermi velocity, e.g., $l_{\rm ee}=v_F \tau_{\rm ee}$. It is easy to check that the conventional parabolic-like (Poiseuille) profile $u_{\parallel}(y) \propto y\left(L-y\right)$ is obtained in the case $\lambda_{\rm G}\gg L$.

As for the electric and chiral charge densities, the solution to Eq.~(\ref{Hydro-eqs-N-steady}) is trivial, $N(y)=0$. The chiral charge density that satisfies Eq.~(\ref{Hydro-eqs-N5-steady}) reads
\begin{equation}
\label{Hydro-cond-N5-sol-i}
N_{5}(y) = -N_{W}e^2\nu_{\rm eff}(\mu) \tau_{\rm ei,5} v_{\Omega} E \left[1 - \frac{\cosh{\left(\frac{L-2y}{2\lambda_{\rm G,5}}\right)}}{\cosh{\left(\frac{L}{2\lambda_{\rm G,5}}\right)}}\right].
\end{equation}
Here, we introduce the chiral Gurzhi length
\begin{equation}
\label{Hydro-cond-lambda-G5}
\lambda_{\rm G,5} = \sqrt{D_{\rm ee} \tau_{\rm ei,5}} = \sqrt{\frac{l_{\rm ee} l_{\rm ei,5}}{3}},
\end{equation}
cf. Eq.~(\ref{Hydro-cond-lambda-G}).

By using the electric current density given in Eq.~(\ref{hydro-eqs-J}), we obtain the following normal $\sigma_0(y)$ and anomalous $\sigma_{\rm anom}(y)$ components of the longitudinal conductivity $\sigma(y)$:
\begin{eqnarray}
\label{Hydro-cond-sigma0-def}
\sigma_0(y)\!\! &=& \!\! \frac{v_F^2 \tau_{\rm eff} N_0^2}{w_0} \left[1 - \frac{\cosh{\left(\frac{L-2y}{2\lambda_{\rm G}}\right)}}{\cosh{\left(\frac{L}{2\lambda_{\rm G}}\right)}}\right] \nonumber\\
&\stackrel{\lambda_{\rm G}\gg L}{\approx}& \frac{5N_0^2 y\left(L-y\right)}{2w_0\tau_{\rm ee}},\\
\label{Hydro-cond-sigma-anom-def}
\sigma_{\rm anom}(y)\!\! &=& \!\!
N_{W} e^2 \nu_{\rm eff}(\mu) v_{\Omega}^2 \tau_{\rm ei,5}\left[1 - \frac{\cosh{\left(\frac{L-2y}{2\lambda_{\rm G,5}}\right)}}{\cosh{\left(\frac{L}{2\lambda_{\rm G,5}}\right)}}\right] \nonumber\\
&\stackrel{\lambda_{\rm G,5}\gg L}{\approx}&
3N_{W} e^2 \nu_{\rm eff}(\mu) \left(\frac{v_{\Omega}}{v_F}\right)^2 \frac{y\left(L-y\right)}{2\tau_{\rm ee}}.
\end{eqnarray}

In order to investigate the Gurzhi effect and the role of the chiral anomaly in it, we calculate the conductivity averaged over the channel width. Let us start with the normal part of the conductivity. We consider two characteristic cases: (i) Ohmic regime $\lambda_{\rm G}\ll L$ with a flat profile of the current and (ii) Poiseuille regime $\lambda_{\rm G}\gg L$ with a parabolic-like profile. The averaged normal part of the conductivity becomes
\begin{eqnarray}
\label{Hydro-cond-sigma0-aver-i}
\lambda_{\rm G}\ll L: \quad \int_0^{L}\frac{dy}{L}\sigma_0(y) &\approx& \frac{v_F^2 \tau_{\rm eff} N_0^2}{w_0} \stackrel{T \ll \mu}{\propto} T^0,\\
\label{Hydro-cond-sigma0-aver-ii}
\lambda_{\rm G}\gg L: \quad \int_0^{L}\frac{dy}{L}\sigma_0(y) &\approx& \frac{5N_0^2 L^2}{12w_0 \tau_{\rm ee}} \stackrel{T \ll \mu}{\propto} T^2.
\end{eqnarray}
Evidently, the normal part of the electric conductivity grows with $T$ for $\lambda_{\rm G}\gg L$. This seemingly counterintuitive behavior is explained by the decrease of momentum-relaxing scattering events with the rise of the electron fluid viscosity $\eta_{\rm dyn} \sim \tau_{\rm ee}$ and is known as the Gurzhi effect~\cite{Gurzhi:1963}.

In the case of the anomalous part of the conductivity given by Eq.~(\ref{Hydro-cond-sigma-anom-def}), we derive
\begin{eqnarray}
\label{Hydro-cond-sigma-anom-aver-i}
\lambda_{\rm G,5}\!\ll \! L: \, \int_0^{L}\!\!\frac{dy}{L}\sigma_{\rm anom}(y) &=&
N_{W} e^2 v_{\Omega}^2 \nu_{\rm eff}(\mu) \tau_{\rm ei,5}\! \stackrel{T \ll \mu}{\propto}\! T^0,\nonumber\\
\\
\label{Hydro-cond-sigma-anom-aver-ii}
\lambda_{\rm G,5}\!\gg \! L: \, \int_0^{L}\!\!\frac{dy}{L}\sigma_{\rm anom}(y) &=& \frac{N_{W} e^2 v_{\Omega}^2 \nu_{\rm eff}(\mu) L^2}{12\tau_{\rm ee}} \!  \stackrel{T \ll \mu}{\propto} \! T^2.\nonumber\\
\end{eqnarray}
Similarly to the electrical conductivity at $B=0$, the anomalous part $\sigma_{\rm anom}$ increases with $T$ at $T\ll \mu$; cf. Eqs.~(\ref{Hydro-cond-sigma0-aver-ii}) and (\ref{Hydro-cond-sigma-anom-aver-ii}). Therefore, in analogy with the Gurzhi effect, we dub this phenomenon the anomalous Gurzhi effect.

The result in Eq.~(\ref{Hydro-cond-sigma-anom-aver-ii}) might look surprising because it does not contain an internode scattering rate but depends on the electron-electron scattering time $\tau_{\rm ee}$. The latter type of scattering does not relax the chiral charge and, consequently, should not be crucial for the anomalous transport. This apparent conundrum can be resolved by noting that frequent electron-electron collisions prevent electrons from reaching the boundaries which are the primary source for chirality relaxation in this regime. This is reminiscent of the conventional Gurzhi effect where the electron fluid relaxes its momentum at the boundaries.

Since $\lambda_{\rm G}\ll \lambda_{G,5}$ (see also Sec.~\ref{sec:Hydro-num} for the estimates), it is possible to realize the mixed \emph{chiral hydrodynamic regime} with $\lambda_{\rm G}\ll L \ll \lambda_{\rm G,5}$. In this case, only the anomalous Gurzhi effect is observed. The main regimes of transport, namely (i) Ohmic ($\lambda_{\rm G}\ll L$ and $\lambda_{\rm G, 5}\ll L$), (ii) chiral hydrodynamic ($\lambda_{\rm G}\ll L \ll \lambda_{\rm G,5}$), and (iii) hydrodynamic ($\lambda_{\rm G}\gg L$ and $\lambda_{\rm G, 5}\gg L$) are schematically shown in Fig.~\ref{fig:MP-cor-phase-diag}.

\subsection{Numerical estimates}
\label{sec:Hydro-num}

Let us compare the magnitudes of the normal $\sigma_0$ and anomalous $\sigma_{\rm anom}$ corrections to the conductivity and estimate them for realistic parameters. In our estimates, we use the parameters $v_{F}\approx 3\times 10^7~\mbox{cm/s}$, $\mu\approx 20~\mbox{meV}$, $\tau_{5}\approx 60~\mbox{ps}$, $\tau \approx 0.38~\mbox{ps}$, $N_{W}=24$, and the dielectric constant $\varepsilon=36$ quoted in Refs.~\cite{Arnold-Felser:2016b,Zhang-Hasan-TaAs:2016,Buckeridge-Sokol:2016} for the Weyl semimetal TaAs. In addition, we assume $T=50~\mbox{K}$. The effective fine-structure constant can be estimated as $\alpha_{\rm eff} = e^2/(\hbar v_F \varepsilon) \approx0.21$.

Let us start with the \emph{Ohmic regime} ($\lambda_{\rm G}\ll L$ and $\lambda_{\rm G,5}\ll L$). By using Eqs.~(\ref{Hydro-cond-sigma0-aver-i}) and (\ref{Hydro-cond-sigma-anom-aver-i}), we obtain
\begin{equation}
\label{hydro-num-sigmasigma-Ohmic}
\frac{\sigma_{\rm anom}}{\sigma_0} = N_{W} \left(\frac{v_{\Omega}}{v_F}\right)^2 \frac{e^2 w_0\nu_{\rm eff}(\mu)}{N_0^2} \frac{\tau_{\rm ei,5}}{\tau_{\rm eff}}
\approx 2~\left(\frac{B}{1~\mbox{T}}\right)^2.
\end{equation}
It is important to notice that there is an upper limit for the magnetic field for which the above estimate is applicable. Indeed, the CKT is valid for nonquantizing magnetic fields, $B\ll B_{\rm uq}=c\mu^2/(2e \hbar v_F^2)\approx3.5~\mbox{T}$. Here, $B_{\rm uq}$ is the magnetic field at which the ultraquantum limit (i.e., only the lowest Landau level is populated) is reached. For $B\sim B_{\rm uq}$, the formation of the Landau levels must be taken into account. The condition $B\ll B_{\rm uq}$ approximately corresponds to $v_{\Omega}\ll v_{F}$ in Eq.~(\ref{hydro-num-sigmasigma-Ohmic}). In a general case of nonspherical Fermi surfaces, the estimate of the upper limit for the magnetic field is given by a semiclassically weak magnetic field $B_{\rm sw}\approx 0.6~\mbox{T}$~\footnote{A semiclassically weak magnetic field corresponds to $\omega_c\tau_{\rm ei} =1$, where $\omega_c=ev_F^2 B/(c\mu)$ is the cyclotron frequency}. The Hall effect contribution to the conductivity can be ignored for $B\lesssim B_{\rm sw}$; see also Ref.~\cite{Lifshitz-Kaganov:1957}. The semiclassically weak magnetic field provides the most conservative estimate for the upper value of the magnetic field for which our approximations hold.

In the case of the \emph{hydrodynamic regime} ($\lambda_{\rm G}\gg L$ and $\lambda_{\rm G,5}\gg L$), we obtain
\begin{equation}
\label{hydro-num-sigmasigma-Hydro}
\frac{\sigma_{\rm anom}}{\sigma_0} =\frac{3N_{W}}{5} \left(\frac{v_{\Omega}}{v_F}\right)^2 \frac{e^2w_0 \nu_{\rm eff}(\mu)}{N_0^2} \approx 7.7\times10^{-3}~\left(\frac{B}{1~\mbox{T}}\right)^2,
\end{equation}
where we employ Eqs.~(\ref{Hydro-cond-sigma0-aver-ii}) and (\ref{Hydro-cond-sigma-anom-aver-ii}). As one can immediately notice by comparing Eqs.~(\ref{hydro-num-sigmasigma-Ohmic}) and (\ref{hydro-num-sigmasigma-Hydro}), the relative anomalous conductivity is $\tau_{\rm eff}/\tau_{\rm ei,5}$ times smaller in the hydrodynamic regime as compared to the Ohmic regime.

Because $\tau_{\rm ei,5}\gg\tau_{\rm eff}$, it can be possible to achieve a \emph{chiral hydrodynamic regime} ($\lambda_{\rm G}\ll L \ll \lambda_{\rm G,5}$). In this case, while the normal part of the conductivity $\sigma_0$ is described by the Drude-like expression (\ref{Hydro-cond-sigma0-aver-i}), the anomalous part is given by the hydrodynamic-like Eq.~(\ref{Hydro-cond-sigma-anom-aver-ii}). The resulting relative conductivity reads as
\begin{equation}
\label{hydro-num-sigmasigma-hybrid}
\frac{\sigma_{\rm anom}}{\sigma_0} = N_{W} \left(\frac{v_{\Omega}}{v_F}\right)^2 \frac{e^2w_0\nu_{\rm eff}(\mu)}{N_0^2} \frac{L^2}{l_{\rm ee}l_{\rm eff}}.
\end{equation}
Since this hybrid regime is valid for $\lambda_{\rm G} \lesssim L \lesssim \lambda_{\rm G,5}$, the relative conductivity (\ref{hydro-num-sigmasigma-hybrid}) takes intermediate values compared to those in Eqs.~(\ref{hydro-num-sigmasigma-Ohmic}) and (\ref{hydro-num-sigmasigma-Hydro}).

Finally, we provide characteristic values for other parameters used in our calculations. In particular, the Fermi wave length is $\lambda_{F}=v_F\hbar/\mu \approx 0.2~\left(1~\mbox{meV}/\mu\right)~\mu\mbox{m}$, the magnetic length is $l_{B}=\sqrt{\hbar c/(eB)} \approx 2.6\times10^{-2}~\sqrt{1~\mbox{T}/B}~\mu\mbox{m}$, the intranode electron-impurity scattering length is $l_{\rm ei}\approx0.1~\mu\mbox{m}$, and the internode electron-impurity scattering length is $l_{\rm ei,5}\approx17.8~\mu\mbox{m}$.

\section{Kinetic approach}
\label{sec:Kinetic}

To address the transport properties in the crossover regime and to support the conclusions reached in Sec.~\ref{sec:Hydro}, let us solve the Boltzmann equation without any assumption regarding the electron-electron collision rate. We use Eq.~(\ref{CKT-equation-kinetic-equation-simpl}) and the collisions integrals given in Sec.~\ref{sec:CKT-coll}.

In what follows, we treat the effects of the magnetic field perturbatively. We start with the case $B=0$ in Sec.~\ref{sec:Kinetic-B=0}. The first-order corrections to the distribution function and the anomalous part of the electric conductivity are calculated in Sec.~\ref{sec:Kinetic-Bnot0}.

\subsection{Vanishing magnetic field}
\label{sec:Kinetic-B=0}

Let us proceed to the case $B=0$. Since there is a preferred direction along the electric field, the distribution function $n_{\alpha,\eta}(y; \mathbf{p})$ is odd with respect to $\mathbf{p}\to -\mathbf{p}$. Therefore, it is convenient to parametrize the distribution function $n_{\alpha,\eta}(y; \mathbf{p})$ as $n_{\alpha,\eta}(y; \mathbf{p}) = -eE \cos{\theta}\, l_{\alpha,\eta}(y; \mathbf{p})$, where $\theta$ is the angle between $\mathbf{E}$ and $\mathbf{p}$. In the absence of the magnetic field and boundaries, $l_{\alpha,\eta}(y; \mathbf{p})$ corresponds to the mean free path of electrons in conductors.

The electric current density is determined by the function $l_{\alpha,\eta}(y;\mathbf{p})$ summed over all Weyl nodes and averaged over angles, i.e.,
\begin{equation}
\label{Kinetic-B=0-jx}
J_{\parallel}(y) = -e^2E v_F\sum_{\eta} \int_0^{\infty} \frac{dp\,p^2}{(2\pi \hbar)^3} \frac{4\pi}{3} \tilde{l}_{\eta}(y;p) f_{\eta}^{\prime}(p);
\end{equation}
see Eq.~(\ref{CKT-cc-j}) for the definition of current. Here,
\begin{equation}
\label{Kinetic-B=0-tilde}
\tilde{l}_{\eta}(y;p) = \frac{3}{4\pi}\int_{0}^{2\pi} d\varphi \int_{0}^{\pi} d\theta \sin{\theta} \cos^2{\theta}\, l_{\eta}(y; p, \theta,\varphi)
\end{equation}
is the function $\cos^2{\theta}\, l_{\eta}(y; \mathbf{p}) = \cos^2{\theta} \sum_{\alpha}^{N_{W}}l_{\alpha,\eta}(y;\mathbf{p})$
averaged over angles. By introducing the following notations,
\begin{eqnarray}
\label{Kinetic-B=0-tilde-underbar}
\underline{\tilde{l}}(y) &=& \frac{\sum_{\eta} \eta \int_0^{\infty} dp\, p^3 \tilde{l}_{\eta}(y; p)\, f_{\eta}^{\prime}(p)}{N_{W}\sum_{\eta} \eta \int_0^{\infty} dp\, p^4 f_{\eta}^{\prime}(p)},\\
\label{Kinetic-B=0-ltot}
l_{\rm tot} &=& v_F\left(\frac{1}{\tau_{\rm ei}} + \frac{1}{2\tau_{\rm ei,5}} +\frac{1}{\tau_{\rm ee}}\right)^{-1},
\end{eqnarray}
we rewrite the kinetic equation (\ref{CKT-equation-kinetic-equation-simpl}) at $B=0$ as
\begin{equation}
\label{Kinetic-B=0-eq-2}
\sin{\theta} \sin{\varphi} \,\partial_y l_{\eta}(y;\mathbf{p}) +\frac{\eta}{l_{\rm tot}}l_{\eta}(y;\mathbf{p})
-N_{W}\frac{\eta p}{l_{\rm ee}}\,\underline{\tilde{l}}(y) = N_{W}.
\end{equation}
The general form of the solution to the above equation with the boundary conditions (\ref{CKT-BC-y=0}) and (\ref{CKT-BC-y=L}) is given as an integral equation
\begin{equation}
\label{Kinetic-B=0-tilde-l-1}
\tilde{l}_{\eta}(y;p) = \eta N_{W}\int_{0}^{L}dy^{\prime} K_1\left(\frac{|y-y^{\prime}|}{l_{\rm tot}}\right)  \left[1+\frac{\eta p}{l_{\rm ee}} \underline{\tilde{l}}(y^{\prime})\right],
\end{equation}
where we introduce the following function,
\begin{equation}
\label{Solution-B0-K-1-def}
K_1\left(\frac{|y-y^{\prime}|}{l_{\rm tot}}\right) = \frac{3}{4\pi}\int_{0}^{\pi}d\varphi \int_{0}^{\pi} d\theta \frac{\cos^2{\theta}}{\sin{\varphi}} e^{-\frac{|y-y^{\prime}|}{l_{\rm tot} \sin{\varphi} \sin{\theta}}}.
\end{equation}
The details of the derivation and the explicit form of the solution used in the calculations are given in Appendix~\ref{sec:app-kinetic}.

\begin{figure}[t]
\centering
\includegraphics[width=0.45\textwidth]{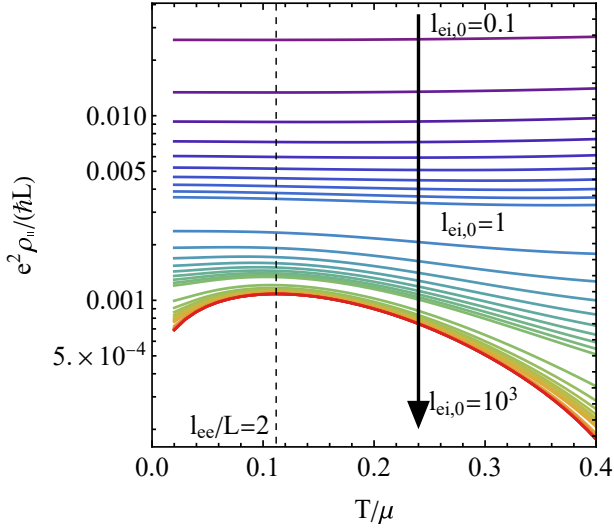}
\caption{
Dependence of the resistivity $\rho_{\parallel}$ on temperature for a few values of the dimensionless impurity scattering length $l_{\rm ei,0}=l_{\rm ei}(p_F)/L$ at $\alpha_{\rm eff}=0.2$, $\lambda_{F}=0.001\, L$, and $N_{W}=24$. A rather small Fermi length is used to emphasize the features of the result. The dashed vertical line corresponds to the temperature at which $l_{\rm ee}=2\,L$.
}
\label{fig:Kinetic-B=0-3D}
\end{figure}

The conductivity averaged over the film width follows from Eq.~(\ref{Kinetic-B=0-jx}),
\begin{equation}
\label{Kinetic-B=0-sigma-def}
\sigma_{\parallel} = -\frac{e^2 v_F}{6\pi^2 \hbar^3}\sum_{\eta} \int_0^{L} \frac{dy}{L} \int_0^{\infty}dp\,p^2 \tilde{l}_{\eta}(y;p) f_{\eta}^{\prime}(p).
\end{equation}
The corresponding resistivity is defined as $\rho_{\parallel}=1/\sigma_{\parallel}$. Numerical results for the resistivity as a function of temperature are shown in Fig.~\ref{fig:Kinetic-B=0-3D} for several values of the electron-impurity scattering length $l_{\rm ei}(p_F)$. The Knudsen peak is clearly visible for $l_{\rm ee}\approx 2\, L$ if $l_{\rm ee}\lesssim L, l_{\rm ei}$. The peak gradually disappears with decreasing electron-impurity scattering length. In the case of strong electron-impurity scattering, $l_{\rm ei}\ll L, l_{\rm ee}$, a temperature-independent diffusive resistivity is obtained.

\subsection{Role of the chiral anomaly}
\label{sec:Kinetic-Bnot0}

In this section, we address the effects of the chiral anomaly on the electric conductivity. According to Eq.~(\ref{CKT-cc-j}), the anomalous part of the electric current is even in the magnetic field and reads
\begin{eqnarray}
\label{Kinetic-Bnot0-j-anom}
&&\mathbf{J}_{\rm anom}(y) = -\frac{e^2}{c}\sum_{\eta}\eta \sum_{\alpha}^{N_{W}} \int \frac{d^3p}{(2\pi \hbar)^3} \left(\mathbf{v}\cdot\mathbf{\Omega}_{\alpha,\eta}\right) \mathbf{B} f_{\eta}^{\prime}(p)\nonumber\\
&&\times n_{\alpha,\eta}^{(1)}(y;\mathbf{p})
= \frac{e^2 v_F}{4\pi^2 \hbar l_{B}^2} \hat{\mathbf{B}} E \sum_{\eta}\eta \int_{\Lambda_{\rm IR}}^{\infty} dp\, \widetilde{\widetilde{l_{5,\eta}^{(1)}}}(y;p) f_{\eta}^{\prime}(p). \nonumber\\
\end{eqnarray}
In the last equation, we use the explicit expression for the Berry curvature and define the function
\begin{equation}
\label{Kinetic-Bnot0-3bar-l}
\widetilde{\widetilde{l_{5,\eta}^{(1)}}}(y; p) = \frac{1}{4\pi} \int_{0}^{2\pi}d\varphi \int_{0}^{\pi} d\theta \sin{\theta}\, l_{5,\eta}^{(1)}(y; p, \theta, \varphi),
\end{equation}
where $l_{5,\eta}^{(1)}(y; \mathbf{p}) = -\sum_{\alpha}^{N_{W}}\chi_{\alpha}n_{\alpha,\eta}^{(1)}(y;\mathbf{p})/(eE)$. Notice that the infrared cutoff $\Lambda_{\rm IR}\simeq \hbar/l_{B}$ is introduced to regularize the momentum integral in Eq.~(\ref{Kinetic-Bnot0-j-anom}). It originates from the fact the CKT breaks down at small momenta where $e\left(\bm{\Omega}_{\alpha,\eta}\cdot \mathbf{B}\right)/c \gtrsim 1$~\cite{Stephanov:2012}; see also Ref.~\cite{Zyuzin:2017}. The cutoff is set to zero in convergent terms.

The function $l_{5,\eta}^{(1)}(y; \mathbf{p})$ satisfies the following kinetic equation,
\begin{eqnarray}
\label{Kinetic-Bnot0-l5-equation}
&&\eta \sin{\theta} \sin{\varphi} \,\partial_y  l_{5,\eta}^{(1)}(y; \mathbf{p})
+\frac{l_{5,\eta}^{(1)}(y; \mathbf{p})}{l_{\rm tot}} -\frac{\left\langle l_{5}^{(1)}(y)\right\rangle}{l_{\rm ee}}\nonumber\\
&&-\left(\frac{1}{l_{\rm ei}} -\frac{1}{2l_{\rm ei,5}}\right) \overline{l_{5,\eta}^{(1)}}(y; \mathbf{p})
= -\frac{N_{W}}{2} \frac{\lambda_{F}^2}{l_{B}^2} \frac{p_F^2}{p^2},
\end{eqnarray}
where the term on the right-hand side in Eq.~(\ref{Kinetic-Bnot0-l5-equation}) originates from the chiral anomaly.

The general form of the solution to Eq.~(\ref{Kinetic-Bnot0-l5-equation}) averaged over the angles reads
\begin{eqnarray}
\label{Kinetic-Bnot0-sol-3bar-0}
&&\widetilde{\widetilde{l_{5,\eta}^{(1)}}}(y; p) = \int_{0}^{L}dy^{\prime} K_2\left(\frac{|y-y^{\prime}|}{l_{\rm tot}}\right)
\Bigg[\!
-\frac{N_{W}}{2} \frac{\lambda_{F}^2}{l_{B}^2} \frac{p_F^2}{p^2} \nonumber\\
&&+\left(\frac{1}{l_{\rm ei}} -\frac{1}{2l_{\rm ei,5}}\right) \widetilde{\widetilde{l_{5,\eta}^{(1)}}}(y^{\prime}; p)
+ \left\langle l_{5}^{(1)}(y^{\prime})\right\rangle \!
\Bigg].
\end{eqnarray}
Here, we define
\begin{equation}
\label{Solution-Bnot0-K-2-def}
K_2\left(\frac{|y-y^{\prime}|}{l_{\rm tot}}\right) = \frac{1}{4\pi}\int_{0}^{\pi}d\varphi \int_{0}^{\pi} d\theta \frac{1}{\sin{\varphi}} e^{-\frac{|y-y^{\prime}|}{l_{\rm tot} \sin{\varphi} \sin{\theta}}}.
\end{equation}

\begin{figure*}[t]
\centering
\subfigure[]{\includegraphics[width=0.45\textwidth]{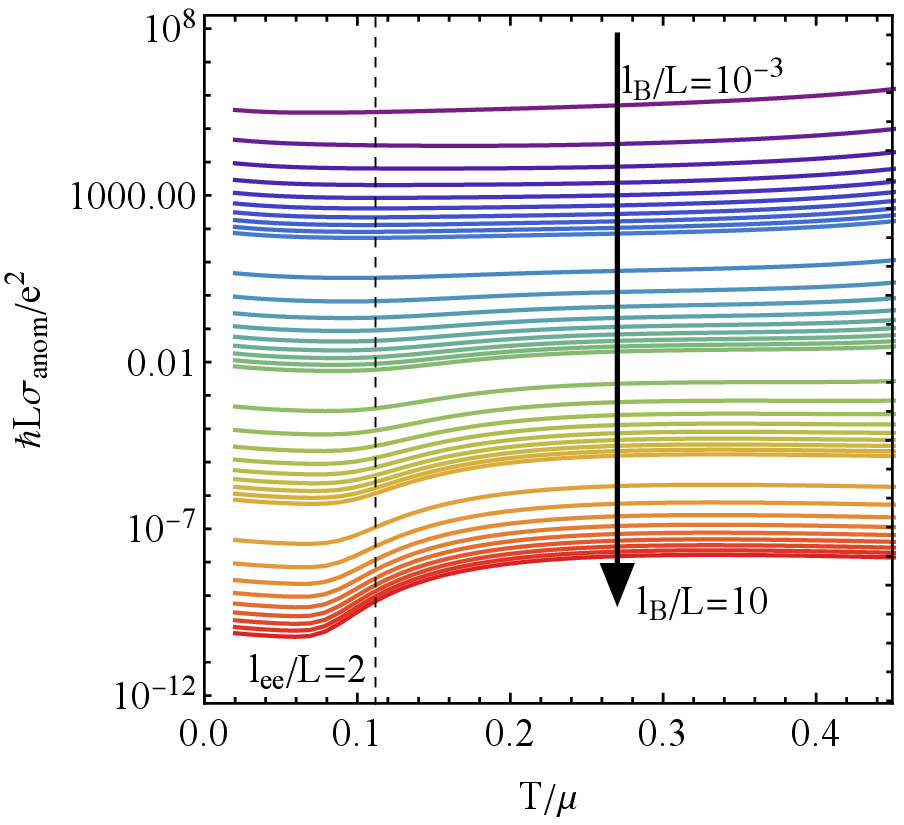}}
\hspace{0.05\textwidth}
\subfigure[]{\includegraphics[width=0.45\textwidth]{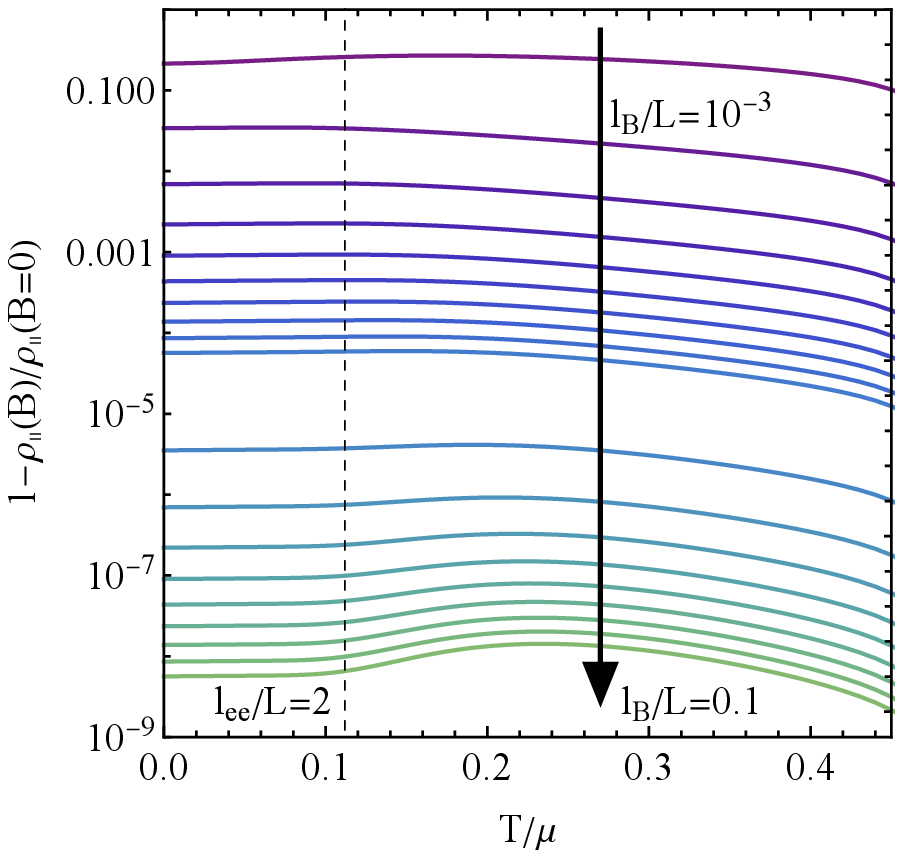}}
\caption{
Dependence of the anomalous part of the conductivity $\sigma_{\rm anom}$ [panel (a)] and the relative anomalous contribution to the resistivity $1-\rho_{\parallel}(B)/\rho_{\parallel}(B=0)$ [panel (b)] on temperature for several values of the magnetic length $l_B=\sqrt{\hbar c/(eB)}$. In all panels, we set $\lambda_{F}=0.001\, L$, $l_{\rm ei,0}=100$, $l_{\rm ei,5}/l_{\rm ei}=100$, $\alpha_{\rm eff}=0.2$, and $N_{W}=24$. The dashed vertical line corresponds to the temperature at which $l_{\rm ee}=2\,L$.
}
\label{fig:Kinetic-Bnot0-sigma-drho}
\end{figure*}

We present the anomalous part of the conductivity $\sigma_{\rm anom}$ averaged over the film width and the relative anomalous contribution to the resistivity $1-\rho_{\parallel}(B)/\rho_{\parallel}(B=0)$ in Figs.~\ref{fig:Kinetic-Bnot0-sigma-drho}(a) and \ref{fig:Kinetic-Bnot0-sigma-drho}(b), respectively. The conductivity $\sigma_{\rm anom}(B)$ follows from Eq.~(\ref{Kinetic-Bnot0-j-anom}) where the solution (\ref{Kinetic-Bnot0-sol-3bar-0}) is used. Evidently, the anomalous part of the electric conductivity demonstrates a nontrivial dependence on temperature and magnetic field. For small magnetic fields, $l_{B}\gtrsim L$, the conductivity decreases for small temperatures, reaches a local minimum, and then grows again until a local maximum is reached. This dependence originates from the first term in Eq.~(\ref{Kinetic-Bnot0-sol-3bar-0}). The local maximum gradually disappears for large magnetic fields $l_{B}\lesssim L$, where a growth of the anomalous conductivity at $l_{\rm ee}\lesssim L$ is observed, which can be expected from the hydrodynamic analysis.

Therefore, while the calculations in the kinetic theory confirm that the anomalous conductivity can increase with temperature, see Sec.~\ref{sec:Hydro-cond} for the analysis in the hydrodynamic regime, a few interesting features might be realized in the crossover regime of transport. In particular, the nonmonotonic behavior of $\sigma_{\rm anom}$ might appear even in crossover and diffusive regimes. This nontrivial dependence on temperature and the magnetic field is related to the cutoff-dependent term in the current (see also the first term in Eq.~(\ref{Kinetic-Bnot0-sol-3bar-0})).

\section{Summary and discussion}
\label{sec:Summary}

In this work, we investigate the role of the chiral anomaly in hydrodynamic and crossover regimes of transport in a film of Dirac or Weyl semimetals. In the presence of an external static magnetic field applied parallel to the surface of the film, it is found that the anomalous part of the longitudinal resistivity with respect to the magnetic field can demonstrate an analog of the Gurzhi effect. In this case, the resistivity decreases with temperature if electron-electron collisions dominate. This effect is dubbed the anomalous Gurzhi effect.

By using the chiral kinetic theory, we develop hydrodynamic and kinetic approaches that include the effects of chiral anomaly and electron-electron scattering; see Secs.~\ref{sec:Hydro} and \ref{sec:Kinetic}. Amended with appropriate collision integrals as well as diffusive boundary conditions, the kinetic approach is able to describe the transport properties of the film in different transport regimes determined by electron-electron $l_{\rm ee}$, intranode $l_{\rm ei}$, and internode $l_{\rm ei,5}$ electron-impurity scattering lengths.

It is shown that the anomalous part of the conductivity $\sigma_{\rm anom}=\sigma(B)-\sigma(B=0)$ is positive and demonstrates an unexpected scaling with electron-electron and electron-impurity scattering times; see Sec.~\ref{sec:Hydro} for details. Instead of the conventional linear dependence on the internode scattering length $\sigma_{\rm anom} \propto l_{\rm ei,5}$, the anomalous part of the conductivity is inversely proportional to $l_{\rm ee}$, i.e., $\sigma_{\rm anom}\sim L^2/l_{\rm ee}$, if electron-electron scattering dominates. Since $l_{\rm ee}\sim 1/T^2$, the anomalous part of the conductivity can grow with temperature.

The nontrivial scaling of the anomalous part of the conductivity has a similar origin to that in the conventional Gurzhi effect, where the electron fluid relaxes its momentum at the boundaries. Naturally, the internode chirality-mixing scattering that determines the anomalous part of the conductivity takes place at diffusive surfaces in the anomalous Gurzhi effect. In order to reach the surfaces, electrons in the bulk move diffusively with the characteristic length scale determined by electron-electron collisions $\sim L^2/l_{\rm ee}$. This is in stark contrast to the disorder-dominated regime where the characteristic length scale is determined by the internode scattering in the bulk, $l_{\rm ei,5}$. No anomalous Gurzhi effect appears in the latter case. Since the internode scattering length is typically much larger than the intranode one~\cite{Zhang-Hasan-TaAs:2016,Jadidi-Drew-TaAs:2019}, we expect that the regime with the anomalous Gurzhi effect $l_{\rm ee}l_{\rm ei,5} \gg L$ should be more accessible than the one with the conventional Gurzhi effect $l_{\rm ee}l_{\rm ei} \gg L$.

In addition to the hydrodynamic regime, the crossover regime of transport is also investigated. In the absence of the magnetic field, we generalize the study of the relativistic Gurzhi effect~\cite{Kashuba-Molenkamp:2018} to the case of 3D materials with a relativistic-like quasiparticle spectrum. We find that the longitudinal resistivity (in the direction along the surface of the film) shows a nonmonotonic dependence on temperature with a characteristic peak in clean materials $l_{\rm ee}\ll l_{\rm ei}$; see Fig.~\ref{fig:Kinetic-B=0-3D}. The position of the peak is determined by the interplay between the electron-electron scattering length and the thickness of the film $L$. No well-pronounced peak occurs in the disorder-dominated (Ohmic) regime $l_{\rm ei} \ll l_{\rm ee}, L$. The anomalous part of the conductivity in the crossover regime is also found to be nontrivial: the shape of the temperature profile depends noticeably on the magnetic field; see Fig.~\ref{fig:Kinetic-Bnot0-sigma-drho}. Unlike the conventional Gurzhi effect, the nonmonotonic dependence of $\sigma_{\rm anom}$ is determined primarily by the Berry curvature rather than the interplay between $l_{\rm ee}$, $l_{\rm ei}$, and $L$. The growth with $T$ predicted in the hydrodynamic regime is observed for sufficiently strong magnetic fields such that $l_{B}\lesssim L$.

Finally, let us discuss the conditions for the experimental observation of the proposed effects. First of all, we notice that the observation of the Gurzhi effect requires sufficiently clean Weyl or Dirac semimetal to reach hydrodynamic or, at least, crossover regimes of transport. The requirements for the anomalous Gurzhi effect are expected to be weaker because $l_{\rm ei,5}\gg l_{\rm ei}$. Furthermore, the anomalous Gurzhi effect relies on the chiral anomaly activated by the external magnetic field $\mathbf{B}$ applied parallel to the surface of the film. Since the magnitude of the effect scales quadratically with $B$, stronger fields facilitate the observation. Our results obtained in the chiral kinetic theory, however, are limited to the case of classically weak magnetic fields or, in the case of spherical Fermi surfaces, nonquantizing fields. The case of quantizing fields requires a separate investigation and will be reported elsewhere. In addition, we notice that the validity range of our results is limited to the case of small temperatures such that the phonon contribution to the collision integral can be neglected. Finally, while most of our results were presented for the model with symmetric well-separated Weyl nodes, the anomalous Gurzhi effect should be observable in other models as long as the chiral anomaly is realized. The order-of-magnitude estimates of the relative conductivities and characteristic scales for realistic parameters are provided in Sec.~\ref{sec:Hydro-num}. While the anomalous contribution to the electric conductivity in the hydrodynamic regime is weaker compared to that in the Ohmic one, the effect might be still observable under favorable conditions.

\begin{acknowledgments}
The authors acknowledge useful communications with L.~I.~Glazman, E.~V.~Gorbar, P.~Matus, I.~A.~Shovkovy, and P.~Sur\'{o}wka. P.O.S. acknowledges support through the Yale Prize Postdoctoral Fellowship in Condensed Matter Theory. B.T. acknowledges financial support through the DFG (SFB 1170), Cluster of Excellence (EXC2147), the High Tech Agenda Bavaria.
\end{acknowledgments}

\onecolumngrid

\appendix

\section{Useful formulas and thermodynamic relations}
\label{sec:app-1}

In this appendix, we provide a few useful formulas and present the explicit expressions for the thermodynamic variables used in the main text. We focus on the case of symmetric Weyl nodes with quasiparticle energies $\eta v_F p$, where $v_F$ is the Fermi velocity, $p$ is the absolute value of quasiparticle momentum, and $\eta=+$ for conduction and $\eta=-$ for valence bands. By making use of the shorthand notation for the equilibrium distribution function $f_{\eta}^{(0)}(p) =1/[e^{(\eta v_F p-\mu)/T}+1]$, where $\mu$ is the chemical potential and $T$ is temperature, it is straightforward to derive the following expressions:
\begin{eqnarray}
\int\frac{d^3p}{(2\pi \hbar)^3} p^{n-2}  f_{\eta}^{(0)}
&=& -\frac{T^{n+1} \Gamma(n+1) }{2\pi^2 \hbar^3 v_F^{n+1}}  \mbox{Li}_{n+1}\left(-e^{\eta \mu/T}\right),
\label{App-1-integral-3a} \\
\int\frac{d^3p}{(2\pi \hbar)^3} p^{n-2} \frac{\partial f_{\eta}^{(0)}}{\partial p}
&=& \frac{T^{n} \Gamma(n+1)}{2\pi^2 \hbar^3 v_F^{n}}  \mbox{Li}_{n}\left(-e^{\eta \mu/T}\right).
\label{App-1-integral-3b}
\end{eqnarray}
Here, $\Gamma(n)$ is the gamma function and $\mbox{Li}_{n}(x)$ is the polylogarithm function~\cite{Bateman-Erdelyi:book-t1}. The following identities for the polylogarithm functions are useful when taking into account the contributions from valence and conduction bands:
\begin{eqnarray}
\mbox{Li}_{1}\left(-e^{x}\right)-\mbox{Li}_{1}\left(-e^{-x}\right) &=& -x,\\
\mbox{Li}_{2} (-e^{x}) +\mbox{Li}_{2} (-e^{-x})   &=& -\frac{1}{2}\left(x^2+\frac{\pi^2}{3}\right),\\
\mbox{Li}_{3} (-e^{x}) - \mbox{Li}_{3} (-e^{-x})   &=&  -\frac{x}{6}\left(x^2+\pi^2\right),\\
\mbox{Li}_{4} (-e^{x}) + \mbox{Li}_{4} (-e^{-x})   &=&  -\frac{1}{24} \left(x^4 +2\pi^2 x^2 +\frac{7}{15} \pi^4\right).
\end{eqnarray}

Let us proceed to the thermodynamic variables. In terms of the distribution function $f_{\eta}^{(0)}(p)$, the electric charge and energy densities read
\begin{eqnarray}
\label{App-1-def-n}
N_0 &=& -e\sum_{\alpha}^{N_{W}}\sum_{\eta}\eta \int\frac{d^3p}{(2\pi \hbar)^3} f_{\eta}^{(0)}(p) =eN_{W}\frac{T^{3}}{\pi^2 \hbar^3 v_F^{3}}  \left[\mbox{Li}_{3}\left(-e^{\mu/T}\right) -\mbox{Li}_{3}\left(-e^{-\mu/T}\right)\right] = -eN_{W}\frac{\mu \left(\mu^2 +\pi^2T^2\right)}{6\pi^2 v_F^3 \hbar^3},\\
\label{App-1-def-eps}
\epsilon_0 &=& \sum_{\alpha}^{N_{W}}\sum_{\eta}\int\frac{d^3p}{(2\pi \hbar)^3} v_F p f_{\eta}^{(0)}(p) =-N_{W}\frac{3T^{4}}{\pi^2 \hbar^3 v_F^{3}}  \left[\mbox{Li}_{4}\left(-e^{\mu/T}\right) +\mbox{Li}_{4}\left(-e^{-\mu/T}\right)\right] \nonumber\\
&=& \frac{N_{W}}{8\pi^2 \hbar^3v_F^3} \left(\mu^4 +2\pi^2T^2\mu^2 +\frac{7\pi^4T^4}{15}\right),
\end{eqnarray}
where $\sum_{\alpha}^{N_{W}}$ denotes the summation over $N_{W}$ Weyl nodes, $\sum_{\eta}$ stands for the summation over bands, and $-e$ is the charge of the electron. Notice that when summing over electrons and holes, one should subtract the contribution of the Dirac sea, i.e.,
\begin{equation}
\label{App-1-eta-sum}
\sum_{\eta=\pm} \eta f^{(0)}_{\eta}(p) = f_{+}^{(0)}(p) -\left[1-f_{-}^{(0)}(p)\right].
\end{equation}

\section{Collision integrals}
\label{sec:app-I-coll}

\subsection{Electron-impurity collision integral}
\label{sec:app-I-coll-ei}

By using the Fermi golden rule (see, e.g., Ref.~\cite{Abrikosov:book-1988}), the electron-impurity collision integral is defined as
\begin{equation}
\label{app-I-coll-def}
I\left[f_{\alpha,\eta}(t,\mathbf{r};\mathbf{p})\right] = - \sum_{\beta}^{N_{W}}\int \frac{d^3p^{\prime}}{(2\pi \hbar)^3} \Theta_{\beta}(\mathbf{p}^{\prime}) \frac{2\pi}{\hbar} \left|A_{\alpha, \beta}(\mathbf{p},\mathbf{p}^{\prime})\right|^2 \delta\left[\tilde{\epsilon}_{\alpha}(\mathbf{p})-\tilde{\epsilon}_{\beta}(\mathbf{p}^{\prime})\right] \left[f_{\alpha,\eta}(t,\mathbf{r};\mathbf{p})-f_{\beta,\eta}(t,\mathbf{r};\mathbf{p}^{\prime})\right],
\end{equation}
where $\left|A_{\alpha, \beta}(\mathbf{p},\mathbf{p}^{\prime})\right|$ is the scattering amplitude between the Weyl nodes $\alpha$ and $\beta$, $\tilde{\epsilon}_{\alpha}(\mathbf{p}) = \epsilon_{\alpha,\eta}-\left(\mathbf{B}\cdot\mathbf{m}_{\alpha,\eta}\right)$ is the total energy dispersion, $\mathbf{m}_{\alpha,\eta}$ is the orbital magnetic moment, and $\Theta_{\alpha,\eta}=\left[1-e \left(\mathbf{B}\cdot \mathbf{\Omega}_{\alpha,\eta}\right)/c\right]$ corresponds to the renormalization of the phase-space volume; see, e.g., Refs.~\cite{Chang-Niu:1996,Son:2013,Xiao-Niu:rev-2010} for details. In these expressions, $\mathbf{\Omega}_{\alpha,\eta}=\mathbf{\Omega}_{\alpha,\eta}(\mathbf{p})$ is the Berry curvature and $c$ is the speed of light. In our studies of the chiral anomaly, we neglect the magnetic moment and the phase-space renormalization, see also Sec.~\ref{sec:CKT-eqs} and Appendix~\ref{sec:app-MP-cor}. Then, the collision integral given in Eq.~(\ref{app-I-coll-def}) becomes
\begin{equation}
\label{CKT-I-coll-1}
I\left[f_{\alpha,\eta}(t,\mathbf{r};\mathbf{p})\right] \approx \sum_{\beta}^{N_{W}}\int \frac{d^3p^{\prime}}{(2\pi \hbar)^3} \frac{2\pi}{\hbar} \left|A_{\alpha, \beta}(\mathbf{p},\mathbf{p}^{\prime})\right|^2 \delta\left[\epsilon_{\alpha}(p)-\epsilon_{\beta}(p^{\prime})\right] \left[n_{\alpha,\eta}(t,\mathbf{r};\mathbf{p}) f_{\alpha,\eta}^{\prime}(p) -n_{\beta,\eta}(t,\mathbf{r};\mathbf{p}^{\prime}) f_{\beta,\eta}^{\prime}(p^{\prime})\right].
\end{equation}
Here, we parametrized the distribution function as
\begin{equation}
\label{app-I-coll-f-chi}
f_{\alpha,\eta}(t,\mathbf{r};\mathbf{p}) =  f_{\alpha,\eta}^{(0)}(p)- f_{\alpha,\eta}^{\prime}(p) n_{\alpha,\eta}(t,\mathbf{r};\mathbf{p}),
\end{equation}
where $f_{\alpha,\eta}^{(0)}(p)$ is the standard Fermi-Dirac distribution function and $f_{\alpha,\eta}^{\prime}(p)$ is its derivative with respect to energy.

We assume a short-range impurity potential $U(\mathbf{r}) = \sum_j u_0 \delta(\mathbf{r}-\mathbf{r}_j)$ and neglect the matrix element $\sim \left(\hat{\mathbf{p}}\cdot \hat{\mathbf{p}}^{\prime}\right)$, which stems from the overlap of the wave functions. In this case $\left|A_{\alpha, \beta}(\mathbf{p},\mathbf{p}^{\prime})\right|^2 \approx n_{\rm imp} u_0^2/2$, where $n_{\rm imp}$ is the density of impurities.

The intranode part $\beta=\alpha$ of the collision integral (\ref{CKT-I-coll-1}) reads
\begin{eqnarray}
\label{app-I-coll-matrix-intra}
I_{\rm intra}\left[f_{\alpha,\eta}(t,\mathbf{r};\mathbf{p})\right] &\approx& \int \frac{d^3p^{\prime}}{(2\pi \hbar)^3} \frac{\pi}{\hbar} n_{\rm imp} u_0^2\, \delta\left[\epsilon_{\alpha}(p)-\epsilon_{\alpha}(p^{\prime})\right] \left[n_{\alpha,\eta}(t,\mathbf{r};\mathbf{p}) f_{\alpha,\eta}^{\prime}(p) -n_{\alpha,\eta}(t,\mathbf{r};\mathbf{p}^{\prime}) f_{\alpha,\eta}^{\prime}(p^{\prime})\right] \nonumber\\
&=&\frac{n_{\alpha,\eta}(t,\mathbf{r};\mathbf{p}) -\overline{n_{\alpha,\eta}}(t,\mathbf{r};p)}{\tau_{\rm ei}(p)} f_{\alpha,\eta}^{\prime}(p).
\end{eqnarray}
Here, the intranode electron-impurity scattering rate is
\begin{equation}
\label{app-I-coll-tau-alpha-local}
\frac{1}{\tau_{\rm ei}(p)} = \frac{\pi}{\hbar} \nu(p) n_{\rm imp}u_0^2,
\end{equation}
the density of states (DOS) for the linear energy spectrum is
\begin{equation}
\label{app-I-coll-DOS-def}
\nu(p)=\frac{p^2}{2\pi^2 \hbar^3 v_F},
\end{equation}
and the averaging over the angles is defined as
\begin{equation}
\label{app-I-coll-bar-local-1}
\overline{n_{\alpha,\eta}}(t,\mathbf{r};p)=
\frac{1}{4\pi}
\int_{0}^{2\pi}d\varphi^{\prime} \int_{0}^{\pi} d\theta^{\prime}\, \sin{\theta^{\prime}} n_{\alpha,\eta}(t,\mathbf{r}; p, \theta^{\prime}, \varphi^{\prime}).
\end{equation}
In the case of the internode scattering, the corresponding part of the collision integral reads as
\begin{eqnarray}
\label{app-I-coll-matrix-inter}
I_{\rm inter}\left[f_{\alpha,\eta}(t,\mathbf{r};\mathbf{p})\right] &\approx& \sum_{\beta \neq\alpha}\int \frac{d^3p^{\prime}}{(2\pi \hbar)^3} \frac{\pi}{\hbar} n_{\rm imp} u_0^2 \delta\left[\epsilon_{\alpha}(p)-\epsilon_{\beta}(p^{\prime})\right] \left[n_{\alpha,\eta}(t,\mathbf{r};\mathbf{p}) f_{\alpha,\eta}^{\prime}(p) -n_{\beta,\eta}(t,\mathbf{r};\mathbf{p}^{\prime}) f_{\beta,\eta}^{\prime}(p^{\prime})\right] \nonumber\\
&=& \sum_{\beta \neq\alpha} \frac{n_{\alpha,\eta}(t,\mathbf{r};\mathbf{p}) -\overline{n_{\beta,\eta}}(t,\mathbf{r};p)}{\tau_{\alpha,\beta}(p)} f_{\alpha,\eta}^{\prime}(p).
\end{eqnarray}

Assuming the model with symmetric well-separated Weyl nodes used in the main text, there is a transfer only between the nodes within the same pair, i.e., $(\alpha,-\alpha)$. Then, by defining $\tau_{\alpha,-\alpha}(p)= 2\tau_{\rm ei,5}(p)$, we rewrite the internode collision integral (\ref{app-I-coll-matrix-inter}) as
\begin{equation}
\label{app-I-coll-matrix-inter-1}
I_{\rm inter}\left[f_{\alpha,\eta}(t,\mathbf{r};\mathbf{p})\right] = \frac{n_{\alpha,\eta}(t,\mathbf{r};\mathbf{p}) -\overline{n_{-\alpha,\eta}}(t,\mathbf{r};p)}{2\tau_{\rm ei,5}(p)} f_{\alpha,\eta}^{\prime}(p).
\end{equation}
We assume that $\tau_{\rm ei,5}(p)$ has the same functional dependence as $\tau_{\rm ei}(p)$ given in Eq.~(\ref{app-I-coll-tau-alpha-local}).

\subsection{Electron-electron collision integral}
\label{sec:app-I-coll-ee}

We describe the electron-electron collisions by using the Callaway ansatz~\cite{Callaway:1959,Jong-Molenkamp:1995}:
\begin{equation}
\label{app-I-coll-ee-I-ee-def}
I_{\rm ee}\left[f_{\alpha,\eta}(t,\mathbf{r};\mathbf{p})\right] = -\frac{f_{\alpha,\eta}(t,\mathbf{r};\mathbf{p}) - f_{\alpha,\eta}^{(\mathbf{u})}(t,\mathbf{r};\mathbf{p})}{\tau_{\rm ee}}.
\end{equation}
Here, $1/\tau_{\rm ee}$ is the electron-electron scattering rate and
\begin{equation}
\label{app-I-coll-ee-fu-def}
f_{\alpha,\eta}^{(\mathbf{u})}(t,\mathbf{r};\mathbf{p}) = \frac{1}{e^{\left[\epsilon_{\alpha,\eta} -\mu -\left(\mathbf{u}(t,\mathbf{r})\cdot\mathbf{p}\right)-\delta\mu_{\alpha}(t,\mathbf{r})\right]/T}+1}
\end{equation}
is the hydrodynamic distribution function with the drift velocity $\mathbf{u}(t,\mathbf{r})$. As we will show below, $\delta\mu_{\alpha}(t,\mathbf{r})$ is needed to conserve the electric charge in each of the Weyl nodes separately. Indeed, the internode scattering usually involves a large momentum transfer. Since the electron-electron collisions conserve momentum, they cannot lead to a transfer between the Weyl nodes and, consequently, change the electric charge density in each of the nodes. It is worth noting that we neglect Umklapp scattering, which does not conserve the total momentum. Finally, we use an electron-electron scattering rate $1/\tau_{\rm ee}$ averaged over momentum; see also Sec.~\ref{sec:CKT-coll}.

In the linear response regime, the deviations from the local equilibrium are weak and one can use the following expansion:
\begin{equation}
\label{app-I-coll-fu-app}
f_{\alpha,\eta}^{(\mathbf{u})}(t,\mathbf{r};\mathbf{p}) \approx f_{\alpha,\eta}^{(0)}(p) - \left[\left(\mathbf{u}(t,\mathbf{r})\cdot\mathbf{p}\right)+\delta\mu_{\alpha}(t,\mathbf{r})\right] f_{\alpha,\eta}^{\prime}(p).
\end{equation}

Integrating over momenta, summing over all bands, and requiring that the electron-electron collision integral conserves the electric charge per Weyl node, we obtain
\begin{equation}
\label{app-I-coll-aver-def}
\delta\mu_{\alpha}(t,\mathbf{r}) = \left\langle n_{\alpha}(t,\mathbf{r})\right\rangle =  \frac{1}{\sum_{\eta}\eta\int d^3p\, f_{\alpha,\eta}^{\prime}(p)} \sum_{\eta}\eta\int d^3p\, n_{\alpha,\eta}(t,\mathbf{r};\mathbf{p}) f_{\alpha,\eta}^{\prime}(p).
\end{equation}

The drift velocity $\mathbf{u}(t,\mathbf{r})$ can be found by demanding the conservation of the total momentum by the collision integral, i.e.,
\begin{equation}
\label{app-I-coll-I-ee-p}
\sum_{\eta=\pm} \eta \sum_{\alpha}^{N_{W}} \int \frac{d^3p}{(2\pi\hbar)^3}\mathbf{p}I_{\rm ee}\left[f_{\alpha,\eta}(t,\mathbf{r};\mathbf{p})\right] =\mathbf{0}.
\end{equation}
By using Eqs.~(\ref{app-I-coll-ee-I-ee-def}), (\ref{app-I-coll-fu-app}), and (\ref{app-I-coll-I-ee-p}), we derive
\begin{equation}
\label{app-I-coll-u-def}
\mathbf{u}(t,\mathbf{r}) = \frac{3}{\sum_{\eta=\pm} \eta \sum_{\alpha}^{N_{W}} \int d^3p\, p^2 f_{\alpha,\eta}^{\prime}(p)} \sum_{\eta=\pm} \eta \sum_{\alpha}^{N_{W}} \int d^3p\, \mathbf{p}\, n_{\alpha,\eta}(t,\mathbf{r};\mathbf{p}) f_{\alpha,\eta}^{\prime}(p).
\end{equation}

The final result for the electron-electron collision integral (\ref{app-I-coll-ee-I-ee-def}) is
\begin{equation}
\label{app-I-coll-I-ee-fin}
I_{\rm ee}\left[f_{\alpha,\eta}(t,\mathbf{r};\mathbf{p})\right] \approx \frac{n_{\alpha,\eta}(t,\mathbf{r};\mathbf{p}) -\left\langle n_{\alpha}(t,\mathbf{r})\right\rangle -\left(\mathbf{p}\cdot\mathbf{u}(t,\mathbf{r})\right)}{\tau_{\rm ee}} f_{\alpha,\eta}^{\prime}(p).
\end{equation}

\section{Hydrodynamic equations}
\label{sec:app-hydro}

This appendix is devoted to the derivation of the hydrodynamic equations used in Sec.~\ref{sec:Hydro}. In the case $\tau_{\rm ee}\ll \tau_{\rm ei}, \tau_{\rm ei,5}$, we assume that the deviations from the ideal hydrodynamic distribution function $f_{\alpha,\eta}^{(\mathbf{u})}(t,\mathbf{r};\mathbf{p})$ are weak, i.e.,
\begin{equation}
\label{app-hydro-f-def}
f_{\alpha,\eta}(t,\mathbf{r};\mathbf{p})\approx f_{\alpha,\eta}^{(\mathbf{u})}(t,\mathbf{r};\mathbf{p}) + \delta f_{\alpha,\eta}(t,\mathbf{r};\mathbf{p}).
\end{equation}
Here, $f_{\alpha,\eta}^{(\mathbf{u})}(t,\mathbf{r};\mathbf{p})$ is given in Eq.~(\ref{app-I-coll-ee-fu-def}) and $\delta f_{\alpha,\eta}(t,\mathbf{r};\mathbf{p})$ quantifies the deviations from it. In the linear response regime, the fluid velocity is small $u\ll v_F$ allowing us to neglect terms quadratic in $u$.

As in the main text, we focus on the model with well-separated pairs of symmetric Weyl nodes; see Sec.~\ref{sec:CKT-lin}. Then, the Boltzmann equation reads
\begin{eqnarray}
\label{app-hydro-kinetic-equation}
&&\partial_t f_{\alpha,\eta}(t,\mathbf{r};\mathbf{p}) +\left\{-e\mathbf{E}-\frac{e}{c}\left[\mathbf{v}\times \mathbf{B}\right] +\frac{e^2}{c}(\mathbf{E}\cdot\mathbf{B})\mathbf{\Omega}_{\alpha,\eta}\right\}\cdot\partial_{\mathbf{p}} f_{\alpha,\eta}(t,\mathbf{r};\mathbf{p}) \nonumber\\
&&+\left\{\mathbf{v} -e\left[\mathbf{E}\times\mathbf{\Omega}_{\alpha,\eta}\right]
-\frac{e}{c}(\mathbf{v}\cdot\mathbf{\Omega}_{\alpha,\eta})\mathbf{B}\right\}\cdot \bm{\nabla} f_{\alpha,\eta}(t,\mathbf{r};\mathbf{p})
=I\left[f_{\alpha,\eta}\right].
\end{eqnarray}
The collision integral is defined in Appendix~\ref{sec:app-I-coll}.

The general approach deriving the hydrodynamic equations is straightforward: one needs to calculate the moments of the Boltzmann equation~\cite{Landau:t6-2013,Landau:t10-1995,Huang:book}. In what follows, we obtain the hydrodynamic equations in the zero- and first-order approximations.

\subsection{Zero-order hydrodynamic equations}
\label{sec:app-hydro-zero}

Let us start with the zero-order hydrodynamic equations. In this case, we neglect $\delta f_{\alpha,\eta}(t,\mathbf{r};\mathbf{p})$ in Eq.~(\ref{app-hydro-f-def}). Then, the continuity relations for charge and energy densities as well as the Euler equation are straightforwardly derived by averaging the Boltzmann equation given in Eq.~(\ref{app-hydro-kinetic-equation}) with the electric charge $-e$, quasiparticle energy $\epsilon_{\alpha,\eta}$, and momentum $\mathbf{p}$; see also the Supplemental Material of Ref.~\cite{Gorbar:2017vph} for details of the derivation of the hydrodynamic equations in Weyl systems.

The charge continuity equation is derived by using the ansatz (\ref{app-hydro-f-def}) with $\delta f_{\alpha,\eta}(t,\mathbf{r};\mathbf{p})=0$ in Eq.~(\ref{app-hydro-kinetic-equation}), summing over the valence and conduction bands $\sum_{\eta}\eta$, and averaging over the momentum $\int d^3p$. The final result reads
\begin{equation}
\label{app-hydro-zero-continuity}
\partial_t N_{\alpha}(t,\mathbf{r}) + \left(\bm{\nabla}\cdot \mathbf{J}_{\alpha}(t,\mathbf{r})\right) = - \chi_{\alpha} \frac{e^3 \left(\mathbf{E}\cdot\mathbf{B}\right)}{4\pi^2 c \hbar^2} -\frac{N_{\alpha}(t,\mathbf{r})-N_{-\alpha}(t,\mathbf{r})}{2\tau_{\rm ei,5}},
\end{equation}
where $N_{\alpha}(t,\mathbf{r})$ is the partial charge density (electric charge density per Weyl node). The partial current density is
\begin{equation}
\label{app-hydro-zero-J}
\mathbf{J}_{\alpha}(t,\mathbf{r}) = N_{0} \mathbf{u}(t,\mathbf{r}) + \chi_{\alpha} \frac{e^2 \mu_{\alpha}(t,\mathbf{r})}{4\pi^2c \hbar^2} \mathbf{B}.
\end{equation}
Here, we used the effective chemical potential $\mu_{\alpha}=\mu+\delta\mu_{\alpha}(t,\mathbf{r})$ that contains the equilibrium value $\mu$ and the nonequilibrium value $\delta\mu_{\alpha}(t,\mathbf{r})$ induced via external perturbations. The subscript $0$ corresponds to the global equilibrium values of thermodynamic parameters; see Eqs.~(\ref{App-1-def-n}) and (\ref{App-1-def-eps}).

In the model with symmetric well-separated Weyl nodes, $\delta\mu_{\alpha}=\delta\mu+\chi_{\alpha}\delta\mu_{5}$ and it is convenient to separate the electric and chiral densities in Eq.~(\ref{app-hydro-zero-continuity}). For this, we sum over all Weyl nodes in Eq.~(\ref{app-hydro-zero-continuity}) without and with multiplying by $\chi_{\alpha}$. The result reads as
\begin{eqnarray}
\label{app-hydro-zero-continuity-N0}
&&\partial_t N(t,\mathbf{r}) + \left(\bm{\nabla}\cdot \mathbf{J}(t,\mathbf{r})\right) = 0,\\
\label{app-hydro-zero-continuity-N5}
&&\partial_t N_{5}(t,\mathbf{r}) + \left(\bm{\nabla}\cdot \mathbf{J}_{5}(t,\mathbf{r})\right) =
-N_{W} e^2 \nu_{\rm eff}(\mu) \left(\mathbf{E}\cdot\mathbf{v}_{\Omega}\right) +\frac{N_{5}(t,\mathbf{r})}{\tau_{\rm ei, 5}},
\end{eqnarray}
respectively. Here, the anomalous velocity $\mathbf{v}_{\Omega}$ is
\begin{equation}
\label{app-hydro-zero-vOmega-def}
\mathbf{v}_{\Omega}= \frac{e}{4\pi^2 c\hbar^2 \nu_{\rm eff}(\mu)} \mathbf{B}
\end{equation}
and we define the effective DOS
\begin{equation}
\label{app-hydro-zero-nu-eff-def}
\nu_{\rm eff}(\mu) = \frac{\mu^2+\pi^2T^2/3}{2\pi^2 \hbar^3 v_F^3}.
\end{equation}
Then, linearizing in deviations, the electric and chiral charge densities are $N(t,\mathbf{r})=N_0 -eN_{W} \nu_{\rm eff}(\mu) \delta\mu(t,\mathbf{r})$ and $N_5(t,\mathbf{r})=-eN_{W} \nu_{\rm eff}(\mu) \delta\mu_5(t,\mathbf{r})$, respectively. The electric and chiral current densities read
\begin{eqnarray}
\label{app-hydro-zero-J0}
\mathbf{J}(t,\mathbf{r}) &=& N_0 \mathbf{u}(t,\mathbf{r}) -N_{5}(t,\mathbf{r}) \mathbf{v}_{\Omega},\\
\label{app-hydro-zero-J5}
\mathbf{J}_{5}(t,\mathbf{r}) &=& -\left[N(t,\mathbf{r})-N_0\right]\mathbf{v}_{\Omega},
\end{eqnarray}
respectively.

Next, we present the Euler equation for the fluid velocity $\mathbf{u}(t,\mathbf{r})$. It is obtained by summing Eq.~(\ref{app-hydro-kinetic-equation}) over valence and conduction bands $\sum_{\eta}\eta$, multiplying it by $\mathbf{p}$, integrating the resulting equation over momenta, and summing over all nodes. The final result becomes
\begin{equation}
\label{app-hydro-zero-Euler}
\frac{w_0}{v_F^2} \partial_t \mathbf{u}(t,\mathbf{r}) -N_0\mathbf{E} -\frac{1}{c} N_0\left[\mathbf{u}(t,\mathbf{r})\times \mathbf{B}\right] + \bm{\nabla} P(t,\mathbf{r}) = -\frac{w_0 \mathbf{u}(t,\mathbf{r})}{v_F^2 \tau_{\rm eff}}.
\end{equation}
Here, $w_0=\epsilon_0+P_0$ is the enthalpy, $P_0=\epsilon_0/3$ is the pressure, $\epsilon_0$ is the energy density, see Appendix~\ref{sec:app-1} for the definitions, and $1/\tau_{\rm eff}=1/\tau_{\rm ei}+1/(2\tau_{\rm ei,5})$.

Finally, the energy continuity equation is obtained by summing Eq.~(\ref{app-hydro-kinetic-equation}) over the bands $\sum_{\eta}\eta$, multiplying by $\epsilon_{\alpha,\eta}=\eta v_F p$, and integrating it over momenta. We obtain the following equation,
\begin{equation}
\label{app-hydro-zero-energy}
\partial_t \epsilon_{\alpha}(t,\mathbf{r}) +w_{\alpha} \left(\bm{\nabla}\cdot\mathbf{u}(t,\mathbf{r})\right)
-\frac{\chi_{\alpha}\nu_{\rm eff}(\mu)}{2} \left(\mathbf{v}_{\Omega} \cdot\bm{\nabla}\right) \left(\mu_{\alpha}^2(t,\mathbf{r}) + \frac{\pi^2T^2}{3}\right) =0,
\end{equation}
where $\epsilon_{\alpha}$ is the energy density per Weyl node (not to be confused with $\epsilon_{\alpha,\eta}$) and $w_{\alpha}$ is the enthalpy density per Weyl node. Since the second term in the above equation is already of the first order in deviations, we can use $w_{\alpha}=w_0/N_{W}$ in Eq.~(\ref{app-hydro-zero-energy}). Summing over all Weyl nodes, we derive
\begin{equation}
\label{app-hydro-zero-energy-sum}
\partial_t \epsilon(t,\mathbf{r})  +\left(\bm{\nabla}\cdot\mathbf{J}^{\epsilon}(t,\mathbf{r})\right) =0,
\end{equation}
where the energy current density is
\begin{equation}
\label{app-hydro-zero-J-eps}
\mathbf{J}^{\epsilon}(t,\mathbf{r}) = w_0\mathbf{u}(t,\mathbf{r}) -\frac{\mu \mathbf{v}_{\Omega}}{e} N_5(t,\mathbf{r}).
\end{equation}

\subsection{First-order hydrodynamic equations}
\label{sec:app-hydro-first}

Let us derive the first-order corrections to the hydrodynamic equations; see Refs.~\cite{Landau:t10-1995,Huang:book} for a general case as well as Ref.~\cite{Narozhny:rev-2019} for the case of graphene. The first-order correction to the distribution function is determined from the Boltzmann equation (\ref{app-hydro-kinetic-equation}) where the collision integral on the right-hand side describes the electron-electron scattering in the relaxation time approximation; i.e., we assume that $I[f_{\alpha,\eta}]\approx -\delta f_{\alpha,\eta}(t,\mathbf{r})/\tau_{\rm ee}$. This approximation is valid in the hydrodynamic regime, i.e., for $\tau_{\rm ee}\ll\tau_{\alpha,\alpha}, \tau_{\alpha,-\alpha}$. The correction $\delta f_{\alpha,\eta}(t,\mathbf{r})$ follows from Eq.~(\ref{app-hydro-kinetic-equation}):
\begin{eqnarray}
\label{app-hydro-first-df}
\delta f_{\alpha,\eta}(t,\mathbf{r};\mathbf{p}) &=& -\tau_{\rm ee} \Big\{\partial_t -e \left(\mathbf{E}\cdot\partial_{\mathbf{p}}\right) -\frac{e}{c}\left[\mathbf{v}\times \mathbf{B}\right]\cdot\partial_{\mathbf{p}} + \frac{e^2}{c} \left(\mathbf{E}\cdot\mathbf{B}\right) \left(\bm{\Omega}_{\alpha,\eta}\cdot\partial_{\mathbf{p}}\right)
+\left(\mathbf{v}\cdot\bm{\nabla}\right)  -e \left[\mathbf{E}\times \bm{\Omega}_{\alpha,\eta}\right]\cdot \bm{\nabla} \nonumber\\
&-&\frac{e}{c} \left(\mathbf{v}\cdot\bm{\Omega}_{\alpha,\eta}\right) \left(\mathbf{B}\cdot\bm{\nabla}\right)
\Big\} f_{\alpha,\eta}^{(\mathbf{u})}(t,\mathbf{r};\mathbf{p}).
\end{eqnarray}
The above result is used to calculate the dissipative corrections in the hydrodynamic regime, such as viscosity.

In calculating the dissipative coefficients, it is convenient to rewrite the time derivative in Eq.~(\ref{app-hydro-first-df}) in terms of the spatial derivatives. For this, we use the zero-order hydrodynamic equations derived in Appendix~\ref{sec:app-hydro-zero} as well as the following expression,
\begin{eqnarray}
\label{app-hydro-first-dtf}
\partial_t f_{\alpha,\eta}^{(\mathbf{u})}(t,\mathbf{r};\mathbf{p}) = -\left(\mathbf{p}\cdot\partial_t \mathbf{u}(t,\mathbf{r})\right) f_{\alpha,\eta}^{\prime}(p)
+(\partial_t N_{\alpha}(t,\mathbf{r})) \partial_{N_{\alpha}} f_{\alpha,\eta}
+(\partial_t \epsilon_{\alpha}(t,\mathbf{r})) \partial_{\epsilon_{\alpha}} f_{\alpha,\eta}.
\end{eqnarray}
Then, we need to substitute the distribution function (\ref{app-hydro-first-df}) into the Boltzmann equation (\ref{app-hydro-kinetic-equation}) and evaluate the corresponding moments. While being straightforward, the derived expressions are bulky. Therefore, we focus only on a few terms that are relevant for longitudinal transport.

Let us first consider the contributions that give rise to the viscosity $\propto \eta_{\rm dyn} \Delta \mathbf{u}(t,\mathbf{r})$. The corresponding dissipative corrections stem from two terms:
\begin{eqnarray}
\label{app-hydro-first-viscosity-1}
&&\tau_{\rm ee}\sum_{\alpha}^{N_{W}}\sum_{\eta} \eta \int\frac{d^3p}{(2\pi \hbar)^3}\mathbf{p}\left(\mathbf{v}\cdot\bm{\nabla}\right) w_{\alpha} \left(\bm{\nabla}\cdot\mathbf{u}(t,\mathbf{r})\right) \partial_{\epsilon_{\alpha}} f_{\alpha,\eta}(p)
\approx \tau_{\rm ee}\sum_{\alpha}^{N_{W}} \frac{w_{0}}{N_{W}} \frac{\bm{\nabla} \left(\bm{\nabla}\cdot\mathbf{u}(t,\mathbf{r})\right)}{3} \partial_{\epsilon_{\alpha}} \sum_{\eta} \int\frac{d^3p}{(2\pi \hbar)^3} v_Fp f_{\alpha,\eta}(p) \nonumber\\
&&= \tau_{\rm ee}w_0\frac{\bm{\nabla}\left(\bm{\nabla}\cdot\mathbf{u}(t,\mathbf{r})\right)}{3}
\end{eqnarray}
and
\begin{eqnarray}
\label{app-hydro-first-viscosity-2}
&&-\tau_{\rm ee}\sum_{\alpha}^{N_{W}} \sum_{\eta} \eta \int\frac{d^3p}{(2\pi \hbar)^3} \mathbf{p}\left(\mathbf{v}\cdot\bm{\nabla}\right) \left(\mathbf{v}\cdot\bm{\nabla}\right) f_{\alpha,\eta}^{(\mathbf{u})}(t,\mathbf{r};\mathbf{p})
=v_F^2\tau_{\rm ee} \frac{2\bm{\nabla}\left(\bm{\nabla}\cdot\mathbf{u}(t,\mathbf{r})\right) +\Delta \mathbf{u}(t,\mathbf{r})}{15} \sum_{\alpha}^{N_{W}}\sum_{\eta} \eta \int\frac{d^3p}{(2\pi \hbar)^3} p f_{\alpha,\eta}^{\prime}(p) \nonumber\\
&&=-\tau_{\rm ee} w_0 \frac{2\bm{\nabla}\left(\bm{\nabla}\cdot\mathbf{u}(t,\mathbf{r})\right) +\Delta \mathbf{u}(t,\mathbf{r})}{5}.
\end{eqnarray}
The dissipative corrections given in Eqs.~(\ref{app-hydro-first-viscosity-1}) and (\ref{app-hydro-first-viscosity-2}) lead to the conventional viscosity terms. The corresponding Navier-Stokes equation reads
\begin{equation}
\label{app-hydro-first-NS}
\frac{w_0}{v_F^2} \partial_t \mathbf{u}(t,\mathbf{r}) -N_0\mathbf{E} -\frac{1}{c} N_0 \left[\mathbf{u}\times \mathbf{B}\right] + \bm{\nabla} P(t,\mathbf{r}) = -\frac{w_0 \mathbf{u}(t,\mathbf{r})}{v_F^2 \tau_{\rm eff}} +\eta_{\rm dyn} \Delta \mathbf{u}(t,\mathbf{r})+\frac{\eta_{\rm dyn}}{3} \bm{\nabla}\left(\bm{\nabla}\cdot\mathbf{u}(t,\mathbf{r})\right).
\end{equation}
Here, $\eta_{\rm dyn} = w_0\tau_{\rm ee}/5$ is the dynamic viscosity.

Next, we calculate dissipative corrections in the continuity relations. We obtain the following correction:
\begin{equation}
\label{app-hydro-first-diffusion}
e\tau_{\rm ee}\sum_{\eta} \eta \int\frac{d^3p}{(2\pi \hbar)^3} \left(\mathbf{v}\cdot\bm{\nabla}\right) \left(\mathbf{v}\cdot\bm{\nabla}\right) f_{\alpha,\eta}^{(\mathbf{u})}(t,\mathbf{r};\mathbf{p})
= e\frac{v_F^2\tau_{\rm ee}}{3} \Delta \sum_{\eta} \eta \int\frac{d^3p}{(2\pi \hbar)^3} f_{\alpha,\eta}^{(\mathbf{u})}(t,\mathbf{r};\mathbf{p})
= -D_{\rm ee} \Delta N_{\alpha}(t,\mathbf{r}),
\end{equation}
where $D_{\rm ee}=v_F^2 \tau_{\rm ee}/3$. The partial current density including the first-order corrections reads
\begin{equation}
\label{app-hydro-first-J}
\mathbf{J}_{\alpha}(t,\mathbf{r}) = N_{0} \mathbf{u}(t,\mathbf{r}) + \chi_{\alpha} e \delta\mu_{\alpha}(t,\mathbf{r}) \nu_{\rm eff}(\mu) \mathbf{v}_{\Omega}
-D_{\rm ee}\bm{\nabla} N_{\alpha}(t,\mathbf{r}).
\end{equation}

\section{Role of magnetization and phase-space volume corrections}
\label{sec:app-MP-cor}

In the main text, we investigate the role of the chiral anomaly neglecting magnetization and phase-space volume renormalization corrections in the chiral kinetic theory~\cite{Xiao-Niu:rev-2010}. In addition, we neglect the second-order corrections in the magnetic field related to the inter-band effects; see Refs.~\cite{Gao-Niu:2014,Gao-Niu:2015} for the field-induced corrections to the Berry curvature and quasiparticle dispersion relation. Their explicit formulation in the case of Weyl semimetals is given in Ref.~\cite{Gorbar:2017cwv}. In this appendix, we estimate the role of these corrections in the transport and compare the corresponding contribution to the conductivity with that from the chiral anomaly.

To estimate the effect of the magnetization and phase-space volume renormalization corrections, we calculate one of the corrections to the Drude part of current density due to the magnetization
\begin{eqnarray}
\label{MP-cor-estim-j}
\mathbf{J}_{\rm corr}(t,\mathbf{r}) &=&
-e^2\tau_{\rm ei} \sum_{\alpha}^{N_{W}} \sum_{\eta} \eta \int \frac{d^3p}{(2\pi \hbar)^3} \mathbf{v} \left(\mathbf{E}\cdot\partial_{\mathbf{p}}\right) \frac{\left(\epsilon_{\alpha,\eta}^{(1)}\right)^2}{2} f^{\prime \prime}_{\eta}(p)
=-e^2\tau_{\rm ei}\sum_{\alpha}^{N_{W}} \sum_{\eta} \eta \int \frac{d^3p}{(2\pi \hbar)^3} \Bigg\{ \mathbf{v} \epsilon_{\alpha,\eta}^{(1)} f^{\prime \prime}_{\eta}(p) \left(\mathbf{E}\cdot\partial_{\mathbf{p}}\right)\epsilon_{\alpha,\eta}^{(1)} \nonumber\\
&+&\mathbf{v} \frac{\left(\epsilon_{\alpha,\eta}^{(1)}\right)^2}{2} \left(\mathbf{E}\cdot\partial_{\mathbf{p}}\right)f^{\prime \prime}_{\eta}(p)
\Bigg\}
= e^2\frac{v_F^2\tau_{\rm ei}}{3} \left(\frac{eB\hbar v_F}{2c}\right)^2 \frac{\mathbf{E}}{2\pi^2 \hbar^3 v_F} \sum_{\alpha}^{N_{W}} \sum_{\eta} \int_{\Lambda_{\rm IR}}^{\infty} \frac{dp}{p} f^{\prime \prime}_{\eta}(p) \nonumber\\
&&-e^2\frac{v_F^2 \tau_{\rm ei}}{3} \frac{\mathbf{E} B^2 + 2\mathbf{B}\left(\mathbf{E}\cdot\mathbf{B}\right)}{20 \pi^2 \hbar^3}\left(\frac{e\hbar v_F}{2c}\right)^2 \sum_{\eta}\eta \int_{0}^{\infty} dp f^{\prime \prime \prime}_{\eta}(p),
\end{eqnarray}
where $\epsilon_{\alpha,\eta}^{(1)} = -\left(\mathbf{B}\cdot\mathbf{m}_{\alpha,\eta}\right) =\eta e v_Fp \left(\mathbf{B}\cdot \bm{\Omega}_{\alpha,\eta}\right)/c$, $\Lambda_{\rm IR}$ is the infra-red cutoff, and we assume a linear dispersion relation with symmetric Weyl nodes. Notice that the first integral is infra-red convergent after summing over the bands $\sum_{\eta}$. The integral over momenta in the second term in the last expression reads
\begin{equation}
\label{MP-cor-estim-j-Drude-corr-int-2}
\int_{0}^{\infty} dp f^{\prime \prime \prime}_{\eta} = \frac{\eta}{v_F^2 T} \frac{1}{4T \cosh^2{\left(\frac{\mu}{2T}\right)}} \tanh{\left(\frac{\mu}{2T}\right)}.
\end{equation}

For $T\ll\mu$, the correction to the longitudinal conductivity due to the magnetic moment and the renormalization of the phase-space volume follows from the first term in the last expression in Eq.~(\ref{MP-cor-estim-j}) and can be estimated as $\sigma_{\rm corr} \sim \sigma_0 \left(v_{\Omega}/v_F\right)^2$. This correction should be compared with the expressions for the anomalous conductivity given in Eqs.~(\ref{Hydro-cond-sigma-anom-aver-i}) and (\ref{Hydro-cond-sigma-anom-aver-ii}).

In the case of the Ohmic regime ($\lambda_{\rm G,5}\ll L$ and $\lambda_{\rm G}\ll L$), we have $\sigma_{\rm corr}/\sigma_{\rm anom} \sim \tau_{\rm ei}/\tau_{\rm ei,5}$, see also Eq.~(\ref{Hydro-cond-sigma-anom-aver-i}). Since $\tau_{\rm ei,5}\gg \tau_{\rm ei}$, the corrections due to the magnetization and the phase-space volume corrections are not important. For the chiral hydrodynamic regime ($\lambda_{\rm G}\ll L\ll\lambda_{\rm G,5}$), we obtain that $\sigma_{\rm corr}/\sigma_{\rm anom} \sim \left(\lambda_{\rm G}/L\right)^2$; see also Eq.~(\ref{Hydro-cond-sigma-anom-aver-ii}). This is a small correction if $\lambda_{\rm G}\ll L$. Finally, in the hydrodynamic regime ($\lambda_{\rm G}\gg L$ and $\lambda_{\rm G,5}\gg L$), the corrections due to the chiral anomaly and the magnetization are comparable $\sigma_{\rm corr}/\sigma_{\rm anom} \sim 1$. While we expect that our qualitative conclusions about the anomalous Gurzhi effect should hold even in this case, the quantitative details might be different.

\section{Solution to the kinetic equations}
\label{sec:app-kinetic}

In this appendix, we discuss the technical details of the solution to the kinetic equations presented in Sec.~\ref{sec:Kinetic}.

\subsection{Vanishing magnetic field}
\label{sec:app-kinetic-B=0}

We start with the case of a vanishing magnetic field. As we discuss in the main text, the general form of the solution to the kinetic equation (\ref{Kinetic-B=0-eq-2}) reads
\begin{equation}
\label{app-kinetic-B=0-tilde-l-1}
\tilde{l}_{\eta}(y;p) = \eta N_{W}\int_{0}^{L}dy^{\prime} K_1\left(\frac{|y-y^{\prime}|}{l_{\rm tot}}\right)  \left[1+\frac{\eta p}{l_{\rm ee}} \underline{\tilde{l}}(y^{\prime})\right],
\end{equation}
where
\begin{eqnarray}
\label{app-kinetic-B=0-tilde}
\tilde{l}_{\eta}(y;p) &=& \frac{3}{4\pi}\int_{0}^{2\pi} d\varphi \int_{0}^{\pi} d\theta \sin{\theta} \cos^2{\theta}\, l_{\eta}(y; p, \theta,\varphi),\\
\label{app-kinetic-B0-K-1-def}
K_1\left(\frac{|y-y^{\prime}|}{l_{\rm tot}}\right) &=& \frac{3}{4\pi}\int_{0}^{\pi}d\varphi \int_{0}^{\pi} d\theta  \frac{\cos^2{\theta}}{\sin{\varphi}} e^{-\frac{|y-y^{\prime}|}{l_{\rm tot} \sin{\varphi} \sin{\theta}}},\\
\label{app-kinetic-B=0-tilde-underbar}
\underline{\tilde{l}}(y) &=& \frac{1}{N_{W}\sum_{\eta} \eta \int_0^{\infty} dp\, p^4 f_{\eta}^{\prime}(p)} \sum_{\eta} \eta \int_0^{\infty} dp\, p^3 \tilde{l}_{\eta}(y; p) f_{\eta}^{\prime}(p),\\
\label{app-kinetic-B=0-ltot}
l_{\rm tot} &=& v_F\left(\frac{1}{\tau_{\rm ei}} + \frac{1}{2\tau_{\rm ei,5}} +\frac{1}{\tau_{\rm ee}}\right)^{-1}.
\end{eqnarray}
All scattering lengths are introduced by multiplying the scattering time with the Fermi velocity, e.g., $l_{\rm ee}=v_F \tau_{\rm ee}$.

To solve the integral equation (\ref{app-kinetic-B=0-tilde-l-1}), we use the following expansions:
\begin{eqnarray}
\label{app-kinetic-B=0-tilde-l-cos}
\tilde{l}_{\eta}(y;p) &=& \sum_{n=0}^{\infty} \tilde{l}_{n, \eta}(p) \cos{\left(\frac{2\pi n y}{L}\right)},\\
\label{app-kinetic-B=0-tilde-l-n}
\tilde{l}_{n,\eta}(p) &=& \frac{2-\delta_{n,0}}{L} \int_0^{L}dy\, \tilde{l}_{\eta}(y;p) \cos{\left(\frac{2\pi n y}{L}\right)},\\
\label{app-kinetic-B=0-tilde-K-nm}
K_{nm}^{(1)}(p) &=& \frac{2}{L} \int_0^{L}dy \int_0^{L}dy^{\prime} K_1\left(\frac{|y-y^{\prime}|}{l_{\rm tot}}\right) \cos{\left(\frac{2\pi n y}{L}\right)}  \cos{\left(\frac{2\pi m y^{\prime}}{L}\right)},
\end{eqnarray}
which rely on a symmetry with respect to the center of the film $y=L/2$; see also Ref.~\cite{Kashuba-Molenkamp:2018}. Integrals over $y$ and $y^{\prime}$ in Eq.~(\ref{app-kinetic-B=0-tilde-K-nm}) can be straightforwardly calculated. We obtain
\begin{eqnarray}
\label{app-kinetic-B=0-tilde-K-00-calc}
K_{00}^{(1)}(p) &=& \frac{3}{2\pi} L\int_{0}^{\pi}d\varphi \int_{0}^{\pi} d\theta \frac{\cos^2{\theta}}{\sin{\varphi}} \frac{2 \left(\xi-1+e^{-\xi}\right)}{\xi^2},\\
\label{app-kinetic-B=0-tilde-K-nn-calc}
K_{nn}^{(1)}(p) &=& \frac{3}{2\pi} L\int_{0}^{\pi}d\varphi \int_{0}^{\pi} d\theta \frac{\cos^2{\theta}}{\sin{\varphi}} \frac{\xi \left[2\xi e^{-\xi} +\xi(\xi-2)+4\pi^2n^2\right]}{\left(\xi^2+4\pi^2n^2\right)^2},\\
\label{app-kinetic-B=0-tilde-K-nm-calc}
K_{nm}^{(1)}(p) &=& \frac{3}{2\pi} L\int_{0}^{\pi}d\varphi \int_{0}^{\pi} d\theta \frac{\cos^2{\theta}}{\sin{\varphi}} \frac{2 \xi^2 \left(e^{-\xi}-1\right)}{\left(\xi^2+4\pi^2n^2\right)\left(\xi^2+4\pi^2m^2\right)},
\end{eqnarray}
where we used the shorthand notation $\xi = L/\left[(l_{\rm tot} \sin{\varphi} \sin{\theta}\right)$.

By using Eqs.~(\ref{app-kinetic-B=0-tilde-l-cos})--(\ref{app-kinetic-B=0-tilde-K-nm}), the integral equation (\ref{app-kinetic-B=0-tilde-l-1}) can be rewritten as
\begin{equation}
\label{app-kinetic-B=0-tilde-l-mat}
\left(\delta_{nm}+\delta_{n0}\delta_{m0}\right) \tilde{l}_{m,\eta}(p) = \eta N_{W}K_{nm}^{(1)}(p) \left(\delta_{m0} +\frac{\eta p}{l_{\rm ee}} \underline{\tilde{l}}_{m}\right).
\end{equation}
By introducing vectors $1_0=\left\{1,0,0,\ldots\right\}$, $\tilde{l}_{\eta}(p)=\left\{\tilde{l}_{0,\eta}(p), \tilde{l}_{1,\eta}(p),\tilde{l}_{2,\eta}(p), \ldots\right\}$, and $\underline{\tilde{l}}=\left\{\underline{\tilde{l}}_{0}, \underline{\tilde{l}}_{1}, \underline{\tilde{l}}_{2}, \ldots\right\}$, as well as defining matrices
\begin{equation}
\label{app-kinetic-B=0-U-def}
\hat{U}_{nm} = \delta_{mn}+\delta_{m0}\delta_{n0}
\end{equation}
and
\begin{equation}
\label{app-kinetic-B=0-Q-def}
\hat{Q}^{(1)} = \hat{U}^{-1} \hat{K},
\end{equation}
we obtain
\begin{equation}
\label{app-kinetic-B=0-tilde-l-mat-1}
\tilde{l}_{\eta}(p) = \eta N_{W} \hat{Q}^{(1)} \left(1_0 +  \frac{\eta p}{l_{\rm ee}} \underline{\tilde{l}}\right).
\end{equation}
By averaging over momentum, see Eq.~(\ref{app-kinetic-B=0-tilde-underbar}), and solving for $\underline{\tilde{l}}$, we obtain
\begin{equation}
\label{app-kinetic-B=0-tilde-underbar-l}
\underline{\tilde{l}} = N_{W} \left(1 - N_{W}\underline{p\hat{Q}^{(1)}}\frac{1}{l_{\rm ee}}\right)^{-1} \underline{\eta \hat{Q}^{(1)}} 1_0.
\end{equation}
The final result for $\tilde{l}_{\eta}(p)$ in Eq.~(\ref{app-kinetic-B=0-tilde-l-1}) reads
\begin{equation}
\label{app-kinetic-B=0-tilde-l-mat-2}
\tilde{l}_{\eta}(p) = \eta N_{W} \hat{Q}^{(1)} \left[1 + N_{W}\frac{\eta p}{l_{\rm ee}}\left(1 - N_{W}\underline{p\hat{Q}^{(1)}}\frac{1}{l_{\rm ee}}\right)^{-1}  \underline{\eta\hat{Q}^{(1)}}\right] 1_0.
\end{equation}
Notice that the conductivity averaged over the film width, which is defined in Eq.~(\ref{Kinetic-B=0-sigma-def}), depends only on the zeroth component of $\tilde{l}_{\eta}(p)$, i.e., $\tilde{l}_{\eta,0}(p)$. In addition, while the matrices formally have infinite dimension in Eq.~(\ref{app-kinetic-B=0-tilde-l-mat-2}), we restrict their dimension to $5\times5$ in our numerical calculations presented in the main text. We have checked that the increase of the matrix dimension does not lead to any noticeable changes in the results for the parameters used in calculations.

\subsection{Nonzero magnetic field}
\label{sec:app-kinetic-Bnot0}

The solution to the kinetic equation at a nonzero magnetic field is obtained along the same lines as in Appendix~\ref{sec:app-kinetic-B=0}. In particular, we solve the integral equation
\begin{equation}
\label{app-kinetic-Bnot0-sol-3bar-0}
\widetilde{\widetilde{l_{5,\eta}^{(1)}}}(y; p) = \int_{0}^{L}dy^{\prime} K_2\left(\frac{|y-y^{\prime}|}{l_{\rm tot}}\right)
\Bigg[
-N_{W}\frac{\lambda_{F}^2}{2l_{B}^2} \frac{p_F^2}{p^2}
+\left(\frac{1}{l_{\rm ei}} -\frac{1}{2l_{\rm ei,5}}\right) \widetilde{\widetilde{l_{5,\eta}^{(1)}}}(y^{\prime}; p)
+ \left\langle l_{5}^{(1)}(y^{\prime})\right\rangle
\Bigg],
\end{equation}
where $p_F=\mu/v_F$, $\lambda_{F}=\hbar/p_F$, $l_{B}=\sqrt{\hbar c/(eB)}$,
\begin{equation}
\label{app-kinetic-Bnot0-3bar-l}
\widetilde{\widetilde{l_{5,\eta}^{(1)}}}(y; p) = \frac{1}{4\pi} \int_{0}^{2\pi}d\varphi \int_{0}^{\pi} d\theta \sin{\theta}\, l_{5,\eta}^{(1)}(y; p, \theta, \varphi)
\end{equation}
and the average $\left\langle \ldots \right\rangle$ is defined in Eq.~(\ref{app-I-coll-aver-def}). In addition, we use similar expansions to those given in Eqs.~(\ref{app-kinetic-B=0-tilde-l-cos})--(\ref{app-kinetic-B=0-tilde-K-nm}) with the replacement
\begin{equation}
\label{app-kinetic-Bnot0-K-2-def}
K_1\left(\frac{|y-y^{\prime}|}{l_{\rm tot}}\right) \to K_2\left(\frac{|y-y^{\prime}|}{l_{\rm tot}}\right) = \frac{1}{4\pi}\int_{0}^{\pi}d\varphi \int_{0}^{\pi} d\theta \frac{1}{\sin{\varphi}} e^{-\frac{|y-y^{\prime}|}{l_{\rm tot} \sin{\varphi} \sin{\theta}}}.
\end{equation}
In the matrix notation, we have
\begin{equation}
\label{app-kinetic-Bnot0-3bar-l-mat-no-ME}
\widetilde{\widetilde{l_{5,\eta}^{(1)}}}(p) = -N_{W}\frac{\lambda_{F}^2}{2l_B^2} \frac{p_F^2}{p^2} \hat{Q}^{(2)} 1_0 +\frac{1}{l_{\rm ee}} \hat{Q}^{(2)} \left\langle l_{5}^{(1)}\right\rangle.
\end{equation}
Here, we define the following matrix:
\begin{equation}
\label{app-kinetic-Bnot0-Q00-def}
\hat{Q}^{(2)} = \left[\hat{U} - \left(\frac{1}{l_{\rm ei}} -\frac{1}{2l_{\rm ei,5}}\right) \hat{K}^{(2)} \right]^{-1} \hat{K}^{(2)}.
\end{equation}
Performing the averaging, we find $\left\langle l_{5}^{(1)}\right\rangle$. Then, the final result for the solution to the integral equation Eq.~(\ref{app-kinetic-Bnot0-sol-3bar-0}) reads
\begin{equation}
\label{app-kinetic-Bnot0-3bar-l-sol}
\widetilde{\widetilde{l_{5,\eta}^{(1)}}}(p) = -\frac{N_W \lambda_{F}^2}{2l_B^2} \Bigg\{
\frac{p_F^2}{p^2} \hat{Q}^{(2)} +\frac{1}{l_{\rm ee}} \hat{Q}^{(2)}
\left[\hat{1} - \frac{1}{l_{\rm ee}} \left\langle \hat{Q}^{(2)}\right\rangle\right]^{-1} \left\langle \frac{p_F^2}{p^2} \hat{Q}^{(2)}\right\rangle
\Bigg\} 1_0.
\end{equation}
For the momentum-independent scattering lengths, $\left\langle \hat{Q}^{(2)}\right\rangle= \hat{Q}^{(2)}$ and
\begin{equation}
\label{app-kinetic-Bnot0-p2-aver}
\left\langle\frac{p_F^2}{p^2}\right\rangle = \frac{1}{1+\pi^2T^2/\mu^2},
\end{equation}
where we used the expressions presented in Appendix~\ref{sec:app-1}.

As in the case of zero order in the magnetic field, see Appendix~\ref{sec:app-kinetic-B=0}, the anomalous part of the conductivity averaged over the film width depends only on the zeroth component of $\widetilde{\widetilde{l_{5,\eta}^{(1)}}}(p)$. In addition, we restrict the dimension of matrices in Eq.~(\ref{app-kinetic-Bnot0-3bar-l-sol}) to $5\times5$.

\bibliography{library-short}
\end{document}